\documentclass[twocolumn,trackchanges]{aastex701}

\usepackage{multirow}
\usepackage{amsmath}   
\usepackage{float}
\usepackage{booktabs}
\usepackage{multirow}
\usepackage{siunitx}
\usepackage{bm}
\newcommand{\bF}{\bm{F}}
\newcommand{\bU}{\bm{U}}
\newcommand{\bP}{\bm{P}}
\newcommand{\bS}{\bm{S}}
\usepackage{array}
\usepackage{booktabs}
\usepackage{tikz}
\usepackage[T1]{fontenc}
\usepackage[utf8]{inputenc}
\newcolumntype{C}{>{\centering\arraybackslash}c}
\usetikzlibrary{
    arrows.meta,
    positioning,
    fit,
    calc,
    backgrounds
}
\usepackage{xcolor}
\usepackage{tikz}
\usetikzlibrary{positioning}

\definecolor{setupaccent}{HTML}{416A8C}
\definecolor{halfaccent}{HTML}{3F7B70}
\definecolor{fullaccent}{HTML}{6D5C8C}
\begin{document}
\shorttitle{FUGUE}
\shortauthors{Płonka \& Janiuk}

\title{FUGUE: A New GPU-accelerated General Relativistic Magnetohydrodynamics Code with Weak-Field Multipole Self-Gravity and Quasi-Stationary Kerr Black Hole Evolution}

\correspondingauthor{Piotr Płonka}
\email[show]{pplonka@cft.edu.pl}

\author[orcid=0009-0007-8455-7883,gname=Piotr,sname=Płonka]{Piotr Płonka}
\affiliation{Center for Theoretical Physics, Polish Academy of Sciences, Al. Lotnik\'ow 32/46, 02-668 Warsaw, Poland}
\email{pplonka@cft.edu.pl}

\author[orcid=0000-0002-1622-3036,gname=Agnieszka,sname=Janiuk]{Agnieszka Janiuk}
\affiliation{Center for Theoretical Physics, Polish Academy of Sciences, Al. Lotnik\'ow 32/46, 02-668 Warsaw, Poland}
\email{agnes@cft.edu.pl}

\begin{abstract}
	We present {FUGUE}, a new GPU-accelerated code for general relativistic magnetohydrodynamics (GRMHD). The code supports an optional quasi-stationary update of the Kerr black hole mass and spin, as well as weak-field multipole self-gravity computed with a spherical harmonic Poisson solver. The calculations using high-order multipoles in the Poisson solver are expensive, so the self-gravity solver is well suited for GPU acceleration. The code uses shock-capturing methods and a second-order Runge--Kutta (RK2) scheme for time integration. We validate the code with hydrodynamic, magnetohydrodynamic, general relativistic, and convergence tests for the self-gravity module. We present an application of the code to a simulation of a self-gravitating, Toomre-unstable, magnetized accretion disk with a mass of \mbox{$M_{\mathrm{disk}}=1\,M_{\odot}$} around a Kerr black hole. In our demonstration model, gravitational instabilities are associated with changes in the mass accretion rate and the Blandford--Znajek luminosity, and may contribute to gamma-ray burst variability.
		{The performance of the code is equivalent to or exceeds that of the current state-of-the-art GRMHD codes, achieving more than 470 million zone-cycles per second on a single NVIDIA GH200 GPU with linear reconstruction.}
\end{abstract}

\keywords{
	\uat{GPU computing}{1969} ---
	\uat{Magnetohydrodynamical simulations}{1966} ---
	\uat{Computational methods}{1965} ---
	\uat{Black hole physics}{159} ---
	\uat{Accretion}{14}
}
\section{Introduction}

General relativistic magnetohydrodynamic simulations are important tools for studying astrophysical phenomena related to relativistic and highly magnetized plasmas. Numerical studies based on such simulations have led to important results, such as clarifying the nature of magnetorotational instability (MRI) and MRI-driven accretion under strong gravity \citep{HARM2003, DeVilliers2003}, establishing the SANE (standard and normal evolution)/MAD (magnetically arrested disk) accretion classification \citep{Narayan2012}, or demonstrating that highly efficient jets can be generated during the MAD state \citep{Tchekhovskoy2011}. Moreover, the results from GRMHD simulations provided a theoretical basis for analyzing observational data for the first image of a black hole \citep{EHT2019ApJ}. A comparison of state-of-the-art GRMHD codes showed that for standard test problems, the results are consistent among different codes \citep{Porth2019}.

Numerical simulations, especially in three dimensions, are computationally expensive. A way to significantly speed up calculations is the use of GPU accelerators. Therefore, several GPU-accelerated
GRMHD
codes have been developed. Examples of codes for stationary spacetimes include {H-AMR} \citep{H-AMR}, {cuHARM} \citep{cuHARM}, {KHARMA} \citep{KHARMA}, and {ECHO} \citep{DelZanna2024}, whereas examples of codes that can also work with dynamical spacetimes include {AsterX} \citep{AsterX}, {GRaM-X} \citep{Shankar2023}, and {AthenaK} \citep{Stone2026}.

The motivation behind developing a new code, which, in contrast to
fixed-background spacetime codes, takes into account the quasi-stationary
evolution of the mass and spin of the black hole, as well as the
self-gravity of the fluid in an effective weak-field approximation,
is to provide a practical tool for three-dimensional simulations of
collapsars and massive self-gravitating accretion disks around spinning
black holes. Modern approaches to numerical studies of collapsars, which are progenitors of long gamma-ray bursts (GRBs), are usually based on three-dimensional GRMHD simulations \citep{Gottlieb2022,2022ApJ...933L...9G,2025ApJ...985L..26I}, which take into account neither the self-gravity of the fluid nor the evolution of the black hole mass and spin.
The simulations of collapsars in dynamical spacetime are robust; however they are typically limited to axisymmetry \citep{2023ApJ...956..100F,2025PhRvD.111l3017S,2026ApJ...996...57S}. Self-gravity can be particularly important in the accretion disks of collapsars, which can be gravitationally unstable \citep{Metzger_2024, Chen_2025, 2026MNRAS.545f1835L}. In the collapsar disk, non-axisymmetric structures may form and become sources of gravitational waves (GWs; \citealt{2007ApJ...658.1173P,2024ApJ...972L...4G,2025PhRvD.111f3048Y}). Hydrodynamic asymmetries can also generate early GWs in core-collapse supernovae \citep{Cusinato2026}.
Potentially, if the disk is cold enough, self-gravity can amplify non-axisymmetric modes and increase GW emission. %
An advantage over the
post-Newtonian studies of self-gravitating unstable disks, is that with our new code %
the self-gravity module is built on top of the evolving Kerr metric, so that we can study the effects of disk fragmentation during the launching of relativistic jets driven by the Blandford--Znajek process \citep{BlandfordZnajek}.

Another important application of our code is the study of black hole spin-down and spin-up. In contrast to the previous studies devoted to spin evolution \citep{2024ApJ...960...82L,2024ApJ...961..212J,2025arXiv250208732I,2025PhRvD.112l3023L}, the code evolves the black hole spin directly in the simulation at every time step. Consequently, both the MAD state \citep{2003PASJ...55L..69N,Tchekhovskoy2011,2024A&A...692A..37N} and the disk respond to the evolving Kerr background spacetime. Thanks to GPU acceleration, we can carry out three-dimensional studies of self-gravitating disks, relativistic jet launching and propagation, and the evolution of black hole spin more efficiently over longer %
simulation runs.

This article is organized as follows. In Sect.~\ref{sec:GRMHD} we present the equations of general relativistic magnetohydrodynamics. In Sect.~\ref{sec:evol}, we present the quasi-stationary black hole evolution. In Sect.~\ref{Sect:SG}, we explain the implementation of weak-field self-gravity. In Sect.~\ref{SEC:Numerical} we describe the numerical scheme of the code. In Sect.~\ref{sec:validation}, we present the validation of the code in tests for special relativistic hydrodynamics, magnetohydrodynamics, GRMHD, and self-gravity. {We compare our code with state-of-the-art codes.} In Sect.~\ref{sect:perf}, we discuss the kernel structure of the GPU code, its performance and scalability tests, the main limitations of our code, and numerical error control. In Sect.~\ref{sect:Toomre} we present the application of our code to simulations of self-gravitating, Toomre-unstable disks around a rotating black hole in a Kerr spacetime updated quasi-stationarily. In Sect.~\ref{sect:Conclusions} we summarize the article and present our conclusions.

\section{General relativistic magnetohydrodynamics}
\label{sec:GRMHD}
The code solves the covariant equations of ideal magnetohydrodynamics in general relativity, namely the continuity equation, the conservation of energy--momentum, and the induction equation, defined as follows:
\begin{equation}
	\begin{array}{c c c}
		\nabla_\mu \left(\rho u^\mu\right) = 0,
		 &
		\nabla_\mu T^{\mu\nu} = 0,
		 &
		\nabla_\mu \left({}^{*}F^{\mu\nu}\right) = 0,
	\end{array}
	\label{eq:GRMHD}
\end{equation}
where $\rho$ is the rest-mass density, $u^\mu $ is the fluid four-velocity, \mbox{$T^{\mu \nu}$} is the total stress--energy tensor of the magnetized fluid, including both fluid and electromagnetic contributions, and \mbox{${}^{*}F^{\mu\nu}$} is the dual Faraday tensor. The ideal-MHD condition is:
\begin{equation}
	u_\nu F^{\mu\nu}=0,
	\label{eq:MHD_1}
\end{equation}
which implies that the electric field vanishes in the comoving frame. Throughout the paper, we use the metric signature \mbox{$(-,+,+,+)$}, geometrized units with \mbox{$G=c=1$} and Heaviside--Lorentz electromagnetic units. All quantities are expressed in units based on the initial black hole mass. The perfect-fluid contribution to the stress--energy tensor is given by:
\begin{equation}
	T^{\mu\nu}_{\rm fluid} = (\rho + u + p) u^\mu u^\nu + p g^{\mu\nu},
	\label{eq:T_fluid}
\end{equation}
where $u$ is the internal energy per unit volume, $p$ is the gas pressure, and $g^{\mu\nu}$ is the metric tensor. The code uses an ideal-gas equation of state:
\begin{equation}
	p=(\gamma_{\rm ad}-1)u,
\end{equation}
where $\gamma_{\rm ad}$ is the adiabatic index. For the electromagnetic field, the stress--energy tensor is given by:
\begin{equation}
	T^{\mu\nu}_{\rm EM} = b^2 u^\mu u^\nu + \frac{1}{2} b^2 g^{\mu\nu} - b^\mu b^\nu,
	\label{eq:T_EM}
\end{equation}
where \mbox{$b^\mu \equiv u_\nu {}^*F^{\nu\mu}$} is the magnetic four-vector, and \mbox{$b^2=b^\mu b_\mu$}. The magnetic four-vector satisfies \mbox{$u_\mu b^\mu=0$}. The total stress-energy tensor is the sum of the fluid and electromagnetic parts Eqs.~\eqref{eq:T_fluid} and~\eqref{eq:T_EM}, given by:
\begin{equation}
	T^{\mu\nu} _{\rm MHD}=(\rho+u+p+b^2)u^\mu u^\nu+
	\left(p+\frac{b^2}{2}\right)g^{\mu\nu}-b^\mu b^\nu.
	\label{eq:MHD}
\end{equation}
The relation between the coordinate-frame magnetic field \mbox{$B^i \equiv {}^{*}F^{it}$} and the magnetic four-vector is given by:
\begin{equation}
	b^t = g_{i\mu}B^i u^\mu,
	\qquad
	b^i = \frac{B^i+b^t u^i}{u^t}.
\end{equation}
For numerical purposes, the code uses the projected four-velocity $\tilde{u}^{i}$, defined as:
\begin{equation}
	\tilde{u}^{i}=\gamma^{i}_{\ \mu}u^{\mu}=u^{i}+\frac{\beta^{i}}{\alpha}\Gamma,
	\label{eq:projected}
\end{equation}
where \mbox{$\gamma^\mu_{\ \nu}=\delta^\mu_{\ \nu}+n^\mu n_\nu$} is the spatial
projection operator, and \mbox{$n_\mu=(-\alpha,0,0,0)$} is the unit normal one-form to
the hypersurface \mbox{$t=\mathrm{const}$}. The corresponding contravariant vector is \mbox{$n^\mu=(1/\alpha,-\beta^i/\alpha)$}. In this expression, \mbox{$\alpha=1/\sqrt{-g^{tt}}$} is the
lapse function, \mbox{$\beta^i=\alpha^2 g^{ti}$} is the shift vector, and
\mbox{$\Gamma=\alpha u^t$} is the Lorentz factor measured by an Eulerian observer
\citep{McKinney2004,HARM2006}.

The code uses Kerr--Schild coordinates, given by the line element:
\begin{equation}
	\begin{aligned}
		ds^2 ={} &
		-\left(1-\frac{2M_{\rm BH}r}{\Sigma}\right)dt^2
		+\frac{4M_{\rm BH}r}{\Sigma}\,dt\,dr
		\\
		         & +\left(1+\frac{2M_{\rm BH}r}{\Sigma}\right)dr^2
		+\Sigma\,d\theta^2
		\\
		         & +\mathcal A_{\rm KS}\,\sin^2\theta\,d\phi^2
		-\frac{4aM_{\rm BH}r\sin^2\theta}{\Sigma}\,dt\,d\phi
		\\
		         & -2a\left(1+\frac{2M_{\rm BH}r}{\Sigma}\right)
		\sin^2\theta\,dr\,d\phi,
	\end{aligned}
	\label{eq:Kerr}
\end{equation}
where $M_{\rm BH}$ is the black hole mass, $a$ is the dimensional Kerr spin parameter, \mbox{$\Sigma \equiv r^2+a^2\cos^2\theta$}, and \mbox{$\mathcal A_{\rm KS} \equiv \Sigma+a^2\left(1+2M_{\rm BH}r/\Sigma\right)\sin^2\theta$}. The corresponding dimensionless spin of the black hole is \mbox{$a_\ast=a/M_{\rm BH}$}. In general, both the black hole mass and the dimensional Kerr spin parameter may evolve with time.

The metric is subsequently transformed to the code coordinates. This transformation introduces a logarithmic radial grid and concentrates resolution toward the equatorial region. The transformation from the Kerr--Schild coordinates \mbox{$x^\mu=(t,r,\theta,\phi)$} to the so-called modified Kerr--Schild coordinates (MKS) \mbox{${x'}^\mu=(t,x_1,x_2,x_3)$} is given by:
\begin{equation}
	r = e^{x_1}, \qquad \theta = \pi x_2 + \frac{1 - h}{2}\sin(2\pi x_2), \qquad \phi=x_3,
	\label{eq:MKS}
\end{equation}
where \mbox{$h \in (0,1]$} is a parameter that adjusts the grid density near the equator.
The initial conditions are specified in Boyer--Lindquist coordinates \citep{BLcor}, then transformed to Kerr--Schild coordinates and finally to modified Kerr--Schild coordinates. For example, the metric components transform as:
\begin{equation}
	g'_{\mu\nu}
	=
	\frac{\partial x^\alpha}{\partial {x'}^\mu}
	\frac{\partial x^\beta}{\partial {x'}^\nu}
	g_{\alpha\beta}.
\end{equation}
After the transformation, the computational grid is uniform in the numerical coordinates. All vector and tensor components are transformed using the Jacobian of the complete coordinate transformation.
\section{Quasi-Stationary Kerr Black Hole Evolution}
\label{sec:evol}
The code includes an optional quasi-stationary evolution of the central Kerr black hole's mass and spin \citep{Janiuk2018,Dominika2021}, by calculating corresponding fluxes through the radial inner boundary, which is located inside the event horizon. The spacetime remains Kerr at each time step, and the full set of Einstein equations is not solved. Instead, we assume that the evolution is quasi-stationary from one time step to the next. The dimensional Kerr spin parameter evolves according to
\begin{equation}
	\frac{d a}{dt}
	=
	\frac{\dot J}{M_{\rm BH}}
	-
	\frac{a}{M_{\rm BH}}\dot E ,
	\label{eq:spinevol_cont}
\end{equation}
where $\dot J$ and $\dot E$ are the angular-momentum and energy fluxes through the inner boundary. The angular momentum flux $\dot J$ is calculated as
\begin{equation}
	\dot{J} \equiv- \int d\theta d\phi\, \sqrt{-g}\, {T^{r}}_\phi.
	\label{eq:spin}
\end{equation}
The corresponding energy accretion rate is calculated as
\begin{equation}
	\dot{E} \equiv \int d\theta d\phi\, \sqrt{-g}\, {T^{r}}_{t},
	\label{eq:edot}
\end{equation}
and is used to update the black hole mass according to
\begin{equation}
	\frac{dM_{\rm BH}}{dt}=\dot E.
	\label{eq:massevol}
\end{equation}
{The use of energy and angular momentum fluxes for black hole spin and mass evolution is based on \citet{Gammie2004}.} To satisfy energy-momentum conservation (Eq.~\ref{eq:GRMHD}) in quasi-stationary evolving spacetime, a source term should be added to the energy conservation, given as follows:
\begin{equation}
	S_t
	=
	\frac{1}{2}\sqrt{-g}\,
	T^{\mu\nu}\,
	\partial_t g_{\mu\nu}.
	\label{eq:S_t}
\end{equation}
The time derivative of the metric is given by:
\begin{equation}
	\partial_t g_{\mu\nu} = \left(
	\frac{\partial g_{\mu\nu}}{\partial M_{\rm BH}}\dot M_{\rm BH}
	+
	\frac{\partial g_{\mu\nu}}{\partial a}\dot a
	\right),
	\label{eq:S_t_der}
\end{equation}
where $\dot M_{\rm BH}$ and $\dot a$ are given by Eqs.~\ref{eq:massevol} and \ref{eq:spinevol_cont}, respectively. In the code, we take into account the full stress--energy tensor, including both the fluid and electromagnetic parts. The latter is essential to capture the Blandford--Znajek process \citep{BlandfordZnajek}, especially the black hole spin-down, through extraction of its angular momentum and rotational energy.

\section{Weak-field self-gravity approximation}
\label{Sect:SG}
Three-dimensional simulations with dynamical spacetime have recently become possible; however, long-duration models are extremely expensive computationally \citep{Hayashi2025}. One commonly used approach to simplify the problem is the conformally flat approximation \citep{Cheong2020}. For instance, in Schwarzschild spacetime, the spatial metric can be written in a conformally flat form in isotropic coordinates, \mbox{$\gamma_{ij}=\psi^4 f_{ij}$}, where $\psi$ is the conformal factor and $f_{ij}$ is the flat spatial metric. Nevertheless, this condition is not generally valid for Kerr spacetime \citep{Garat2000}. One possible way to include self-gravity is to treat it as an effective enclosed-mass metric \citep{Janiuk2023,Plonka2026}. In this approach, neglecting the radial derivatives of the effective parameters, we can obtain an approximately correct radial monopole acceleration. Nevertheless, such a metric treatment does not satisfy Einstein's equations. A common strategy, for instance in tidal disruption event (TDE) simulations, is to treat self-gravity as a metric perturbation with Newtonian scalar potential \citep{Ryu2020,Chan2026}, which gives the correct Newtonian force.

In this paper, we use a different method of taking into account self-gravity, treating it as a scalar potential that generates a four-force on top of the Kerr background. We solve the Poisson equation:
\begin{equation}
	\nabla^2 \Phi = 4\pi \varrho,
	\label{eq:poi}
\end{equation}
where $\nabla^2$ denotes the Laplacian written in spherical coordinates, and the effective gravitational density \mbox{$\varrho = E+S$}, where \mbox{$E=T_{\mu\nu}n^\mu n^\nu$} is the energy density measured by an Eulerian observer, and \mbox{$S=\gamma^{ij}T_{ij}$} is the trace of the spatial stress tensor \citep{Gourgoulhon2007}. The auxiliary spherical coordinates are numerically identified with the spherical Kerr--Schild coordinates after the MKS (Eq.~\ref{eq:MKS}) mapping (for \mbox{$a\neq0$}, this is only computational identification). The source term can be written as follows:
\begin{equation} \varrho=(\rho+u+p+b^2)(2\Gamma^2-1)+2p-2(b^\mu n_\mu)^2.
	\label{effdensity}
\end{equation}
In the cold (\mbox{$u\ll\rho$} and \mbox{$p\ll\rho$}), non-relativistic \mbox{$(\Gamma\simeq1)$}, and weakly magnetized (\mbox{$b^2\ll\rho$}) limit, the effective gravitational density defined above reduces to the rest-mass density, and the standard Newtonian Poisson equation is obtained. Our source term includes relativistically motivated contributions, and in addition to the rest-mass density, it also contains
internal energy, pressure, and magnetic field energy as sources of the gravitational field \citep{Ehlers2005,Oliveira2014,Hwang2016,Noh2019}.

Our treatment is an effective weak-field source-term model and is not a systematic perturbative solution of the Einstein equations. We compute the self-gravity potential using the flat-space Poisson equation (Eq.~\ref{eq:poi}). This approximation is justified only for the weak gravity of the surrounding matter, \mbox{$|\Phi|\ll1$}. The flat-space Poisson operator is used as a practical and numerical approximation. The largest local errors are expected near the black hole, where the dynamics is dominated by the gravitational field of the Kerr black hole and not by the fluid's weak-field gravity. The accuracy of this approximation is discussed in Sect.~\ref{sect:perf}.

A natural way to solve the Poisson equation in spherical coordinates is to use spherical harmonics and a Green's function. We follow the method described in \citet{Couch2013} to avoid the self-potential error, reconstructing the potential on the cell interfaces. We project the effective density onto spherical harmonics:
\begin{equation}
	\varrho_{\ell m}(r) =\int_{0}^{2\pi}\!\!\int_{0}^{\pi}\varrho(r,\theta,\phi)\,Y_{\ell m}^*(\theta,\phi)\,d\Omega,
	\label{eq:project}
\end{equation}
where $\varrho_{\ell m}$ is the radial coefficient of the effective density in the spherical harmonic expansion, $Y_{\ell m}^*$ is the complex conjugate of the spherical harmonic, and \mbox{$d\Omega=\sin\theta\,d\theta\,d\phi$}. On the modified Kerr--Schild (Eq.~\ref{eq:MKS}) grid, the angular integral is computed using the solid angle of each mapped grid cell. Subsequently, we compute the corresponding spherical harmonic component of the gravitational potential, $\Phi_{\ell m}$, using the Green's function method. We evaluate the solution both at cell centers, $r_i$, and at radial cell interfaces, $r_{i+\frac{1}{2}}$, since the interface values are used for the averaging procedure adopted to avoid the self-potential error. The Green's function solution is given by:
\begin{equation}
	\begin{aligned}
		\Phi_{\ell m}(r)
		=
		-\frac{4\pi}{2\ell+1}
		\Bigg[
		 &
		\frac{1}{r^{\ell+1}}
		\int_{r_{\rm in}}^{r}
		\varrho_{\ell m}(r')
		r'^{\ell+2}\,dr'
		\\
		 & +
		r^\ell
		\int_{r}^{r_{\rm out}}
		\varrho_{\ell m}(r')
		r'^{1-\ell}\,dr'
		\Bigg].
	\end{aligned}
\end{equation}
The solution uses only the effective gravitational density inside the computational domain, \mbox{$r_{\rm in}\le r\le r_{\rm out}$}. The black hole is not included in the Poisson source, because its gravity is already described by the Kerr metric (Eq.~\ref{eq:Kerr}). In the quasi-stationary metric update, we calculate the fluxes (Eq.~\ref{eq:spinevol_cont} and Eq.~\ref{eq:massevol}) through the inner boundary. The integral is evaluated on the logarithmic grid using the corresponding quadrature weights. The potential reconstructed at the radial cell interface is calculated as:
\begin{equation}
	\Phi_{i-\frac{1}{2},j,k}
	=
	\sum_{\ell=0}^{\ell_{\max}}
	\sum_{m=-\ell}^{\ell}
	\Phi_{\ell m}\!\left(r_{i-\frac{1}{2}}\right)
	Y_{\ell m}\!\left(
	\theta_j,
	\phi_k
	\right).
\end{equation}
The other radial face, \mbox{$\Phi_{i+\frac{1}{2},j,k}$}, is computed in the same way using \mbox{$r_{i+\frac{1}{2}}$}. The angular face values are computed analogously at fixed $r_i$, using the face-centered angular coordinates \mbox{$\theta_{j\pm1/2}$} and \mbox{$\phi_{k\pm1/2}$}. The final cell-centered value is computed as the average of the six reconstructed face values, given by:
\begin{equation}
	\begin{aligned}
		\Phi_{i,j,k}
		=
		\frac{1}{6}\Big( &
		\Phi_{i-\frac{1}{2},\,j,\,k}
		+
		\Phi_{i+\frac{1}{2},\,j,\,k}
		+
		\Phi_{i,\,j-\frac{1}{2},\,k}
		\\
		                 & +
		\Phi_{i,\,j+\frac{1}{2},\,k}
		+
		\Phi_{i,\,j,\,k-\frac{1}{2}}
		+
		\Phi_{i,\,j,\,k+\frac{1}{2}}
		\Big).
	\end{aligned}
\end{equation}
The gradient of the gravitational potential is computed using a fourth-order finite-difference approximation \citep{Mandal2023} in code coordinates:
\begin{equation}
	\begin{aligned}
		-\nabla_x\Phi
		 & =
		-\frac{4}{3}
		\frac{\Phi_{i+1,j,k}-\Phi_{i-1,j,k}}{2\Delta x}
		\\
		 & \quad
		+
		\frac{1}{3}
		\frac{\Phi_{i+2,j,k}-\Phi_{i-2,j,k}}{4\Delta x}
		+
		\mathcal{O}(\Delta x^4),
		\label{eq:grad}
	\end{aligned}
\end{equation}
and analogously in the two other directions. Although this stencil is fourth-order accurate, the full self-gravity method is only second-order accurate, because of the cell-centered potential reconstruction. %

The implementation of multipole weak-field self-gravity is built on top of the Kerr metric. We treat the self-gravity contribution as a scalar potential that generates a four-force perpendicular to the fluid four-velocity. We do not modify the Kerr metric itself. We therefore prescribe the self-gravity four-force as:
\begin{equation} G^\mu = -\mathcal{I}^{\mu\nu}\nabla_\nu\Phi, \qquad u_\mu G^\mu=0,
	\label{eq:sg_force_full}
\end{equation}
where the passive magnetohydrodynamic inertial response tensor is given by the following equation:
\begin{equation}
	\mathcal{I}^{\mu\nu} = \left(\rho+u+p+b^2\right) \left(g^{\mu\nu}+u^\mu u^\nu\right) - b^\mu b^\nu,
	\label{eq:mhd_inertia_tensor}
\end{equation}
which is obtained by collecting the acceleration-dependent terms in the projected momentum equation (see Appendix~\ref{TENSOR}). The orthogonality of the four-force implies that it does not directly heat or cool the gas in the comoving reference frame. In the quasi-stationary approximation, we neglect the time derivative of the potential when calculating the four-force. The energy–momentum equation takes the form:
\begin{equation}
	\nabla_\mu T^{\mu\nu} = G^\nu.
	\label{eqgrav}
\end{equation}
Our model is an effective approximation of self-gravity, which can provide non-purely monopole acceleration, which is important in amplifying non-axisymmetric structures. Optionally, the code can also use the \mbox{$\ell=1$} coefficients to include an indirect potential associated with the black hole's reflex acceleration
\citep{Michael2010,Rafikov2026}. The dipole terms are used to calculate the acceleration of the black hole. In the black hole frame, we add the opposite uniform acceleration to the gas. However, the model does not include the full response of the metric to the fluid or gravitational waves.

\section{Numerical scheme}
\label{SEC:Numerical}
The numerical scheme of the code is based on the conservative, finite-volume, shock-capturing general relativistic magnetohydrodynamics code HARM \citep{HARM2003,HARM2006}, with an extension for the quasi-stationary update of the Kerr black hole mass and spin used in \citet{Janiuk2018,Dominika2021}. The code solves the GRMHD equations (Eq.~\ref{eq:GRMHD}) in flux-conservative form given by:
\begin{equation}
	\partial_t \bU({\bP})
	+
	\partial_i \bF^i(\bP)
	=
	\bS_{_{\Gamma}}(\bP)
	+
	\bS_{\Phi}(\bP).
	\label{eq:update}
\end{equation}
The geometrical source term $\bS_{\Gamma}$, representing the contribution of the curved spacetime geometry to the conservative evolution equations, is given by:\begin{equation}
	\bS_{{\Gamma}}
	=
	\sqrt{-g}
	\left(
	0,\,
	T^\kappa{}_{\lambda}\Gamma^\lambda{}_{t\kappa},\,
	T^\kappa{}_{\lambda}\Gamma^\lambda{}_{i\kappa},\,
	0
	\right),
\end{equation}
where $\Gamma^\lambda_{\ \mu\nu}$ are the Christoffel symbols, calculated using a finite difference method in modified Kerr--Schild coordinates. The time derivatives are included separately through $S_t$ (Eq.~\ref{eq:S_t}). When the self-gravity module is enabled, the four-force \mbox{$G_\mu=g_{\mu\nu}G^\nu$} (Eq.~\ref{eq:sg_force_full}) defined in
Sect.~\ref{Sect:SG} is added as:
\begin{equation}
	\bS_{\Phi}=\sqrt{-g}\left(0,\,G_t,\,G_i,\,0\right).
	\label{eq:sg_source}
\end{equation}

In the code, we use conserved variables $\bU$ and primitive variables $\bP$. The conserved variables are evolved in time, while the primitive variables are used to compute the fluxes, source terms, characteristic wave speeds, equation of state and other physical quantities used by the numerical scheme. The vector of conserved variables is defined as follows:
\begin{equation}
	\bU\equiv\sqrt{-g}\left(\rho u^t,\,{T^t}_{t}+\rho u^t,\,{T^t}_{j},\,B^i\right).
	\label{eq:cons}
\end{equation}
where the conserved rest-mass contribution is added to the energy variable for numerical stability. The vector of primitive variables is given by:
\begin{equation}
	\bP \equiv  (\rho, u, \tilde{u}^{i}, {B}^i),
\end{equation}
where $\tilde u^i$ is the projected four-velocity (Eq.~\ref{eq:projected}). The primitive variables are used to calculate the physical fluxes through the cell interfaces. The flux vector is defined as:
\begin{equation}
	\bF^{i}
	\equiv
	\sqrt{-g}
	\left(
	\rho u^{i},\,
	{T^{i}}_{t}+\rho u^{i},\,
	{T^{i}}_{j},\,
	b^{k}u^{i}-b^{i}u^{k}
	\right).
\end{equation}
The numerical fluxes between neighboring cells are computed using an approximate Riemann solver. The code uses either the HLLE (Harten--Lax--van Leer--Einfeldt) Riemann solver \citep{Harten,Einfeldt} or the Lax--Friedrichs \citep{Lax,Friedrichs,Rusanov}. The primitive variables are reconstructed at cell interfaces using the third-order piecewise parabolic method (PPM; \citealt{PPM}). Additionally, we implement a piecewise-linear reconstruction method (PLM), with either the monotonized central (\mbox{\textsc{MC}}) or \mbox{MINMOD} (minimum modulus) slope limiter. The reconstruction method is selected by the user. After reconstructing the left and right primitive states, $\bP_L$ and $\bP_R$, at the cell interface, we calculate the HLLE flux:
\begin{equation}
	\bF_{\rm HLLE}
	=
	\frac{
		\lambda_{\mathrm{max}}\bF_L
		+
		\lambda_{\mathrm{min}}\bF_R
		-
		\lambda_{\mathrm{max}}\lambda_{\mathrm{min}}
		\left(
		\bU_R-\bU_L
		\right)}
	{
		\lambda_{\mathrm{max}}+\lambda_{\mathrm{min}}
	},
	\label{eq:HLLE}
\end{equation}
where \mbox{$\lambda_{\mathrm{max}} = \max(0, \lambda_{+,L}, \lambda_{+,R})$}, and
$\lambda_{+,L}$ and $\lambda_{+,R}$ are the right-going characteristic wave speeds
estimated from the left and right reconstructed states, respectively.
Similarly, \mbox{$\lambda_{\mathrm{min}} = -\min(0, \lambda_{-,L}, \lambda_{-,R})$}, where
$\lambda_{-,L}$ and $\lambda_{-,R}$ are the corresponding left-going characteristic
wave speeds. They are defined as \mbox{$\lambda=dx^i/dt$} and are obtained as the two roots of the approximate relativistic fast-magnetosonic dispersion relation \citep{HARM2003}{, given by:}
\begin{equation}
	(u^\mu k_\mu)^2
	-
	c_{\rm ms}^2
	\left[
		(u^\mu k_\mu)^2 + k^\mu k_\mu
		\right]
	=0,
\end{equation}
where $k_\mu$ is the wave-front normal covector, and $c_{\rm ms}$ is the approximate fast-magnetosonic speed, given by:
\begin{equation}
	c_{\rm ms}^2=c_s^2+v_A^2-c_s^2 v_A^2,
\end{equation}
where \mbox{$c_s^2={\gamma_{\rm ad}\, p}/({\rho+u+p})$} is the relativistic speed of sound, and \mbox{$v_A^2={b^2}/({\rho+u+p+b^2})$} is the Alfvén speed. For the more dissipative local Lax--Friedrichs solver, the interface flux is computed as
\begin{equation}
	\label{eq:lax}
	\bF =
	{1\over 2}\left(\bF_L + \bF_R\right)
	-
	{1\over 2}\lambda_{\mathrm{top}}
	\left(\bU_R - \bU_L\right),
\end{equation}
where
\mbox{$\lambda_{\mathrm{top}}=\max\left(\lambda_{\max},\lambda_{\min}\right)$}
is the maximum characteristic wave speed. The conserved variables are then evolved in time using the calculated fluxes.
\subsection{Flux-constrained transport}
For the evolution of the magnetic field, we use flux-constrained transport \citep{Toth2000,HARM2003}. This method
keeps the discrete magnetic field divergence close to round-off error, as
long as the initial magnetic field is divergence-free and the boundary
conditions are consistent:
\begin{equation}
	\partial_i \mathcal{B}^i=0,
	\qquad
	\mathcal{B}^i=\sqrt{-g}B^i.
\end{equation}
We compute the edge-centered electromotive forces (EMFs) from the face-centered numerical fluxes
of the magnetic field components:
\begin{equation}
	\begin{aligned}
		(\mathcal{E}^1)_{i,j,k}
		 & =
		\frac{1}{4}\big[
			F^2(\mathcal{B}^3)_{i,j,k}
			+
			F^2(\mathcal{B}^3)_{i,j,k-1}
		\\
		 & \qquad
			-
			F^3(\mathcal{B}^2)_{i,j,k}
			-
			F^3(\mathcal{B}^2)_{i,j-1,k}
			\big],
	\end{aligned}
\end{equation}

\begin{equation}
	\begin{aligned}
		(\mathcal{E}^2)_{i,j,k}
		 & =
		\frac{1}{4}\big[
			F^3(\mathcal{B}^1)_{i,j,k}
			+
			F^3(\mathcal{B}^1)_{i-1,j,k}
		\\
		 & \qquad
			-
			F^1(\mathcal{B}^3)_{i,j,k}
			-
			F^1(\mathcal{B}^3)_{i,j,k-1}
			\big],
	\end{aligned}
\end{equation}

\begin{equation}
	\begin{aligned}
		(\mathcal{E}^3)_{i,j,k}
		 & =
		\frac{1}{4}\big[
			F^1(\mathcal{B}^2)_{i,j,k}
			+
			F^1(\mathcal{B}^2)_{i,j-1,k}
		\\
		 & \qquad
			-
			F^2(\mathcal{B}^1)_{i,j,k}
			-
			F^2(\mathcal{B}^1)_{i-1,j,k}
			\big].
	\end{aligned}
\end{equation}
The EMFs are then used to correct the magnetic field fluxes:
\begin{equation}
	\begin{aligned}
		F^1(\mathcal{B}^1)_{i,j,k}
		 & =0,
		\\
		F^1(\mathcal{B}^2)_{i,j,k}
		 & =
		\frac{1}{2}
		\left[
			(\mathcal{E}^3)_{i,j,k}
			+
			(\mathcal{E}^3)_{i,j+1,k}
			\right],
		\\
		F^1(\mathcal{B}^3)_{i,j,k}
		 & =
		-\frac{1}{2}
		\left[
			(\mathcal{E}^2)_{i,j,k}
			+
			(\mathcal{E}^2)_{i,j,k+1}
			\right].
	\end{aligned}
\end{equation}
The fluxes in the other spatial directions are corrected analogously.
After this correction, the finite-volume update of the magnetic field
is equivalent to a discrete curl of the EMFs. This construction uses the same EMF for neighboring faces, so the discrete divergence of the magnetic field remains zero.
\subsection{Conserved-variable update}
The timestep is estimated from the characteristic speeds
computed during the flux calculation. In each direction, the timestep limit is given by:
\begin{equation}
	\Delta t_i
	=
	{\rm CFL}
	\frac{\Delta x_i}
	{\max\left(
		\lambda_{\rm max},
		\lambda_{\rm min}
		\right)},
	\label{eq:timestep}
\end{equation}
where ${\rm CFL}$ is the Courant--Friedrichs--Lewy (CFL) factor, and
$\lambda_{\max}$ and $\lambda_{\min}$ are the characteristic wave speeds defined above. In three dimensions,
the directional limits are combined as
\begin{equation}
	\Delta t
	=
	\left(
	\frac{1}{\Delta t_1}
	+
	\frac{1}{\Delta t_2}
	+
	\frac{1}{\Delta t_3}
	\right)^{-1}.
\end{equation}
The timestep is additionally limited by the safety factor (\mbox{$f_{\rm safe}=0.9$}), the maximum timestep (\mbox{$\Delta t_{\max}=10$}), and the maximum growth factor (\mbox{$f_{\rm grow}=1.3$}).

The conserved variables (Eq.~\ref{eq:update}) are evolved using the second-order explicit midpoint Runge--Kutta method (RK2). We first define the evolution operator:
\begin{equation}
	\mathcal{L}(\bU)
	=
	-\partial_i \bF^i\bigl(\bP(\bU)\bigr)
	+
	\bS_{\Gamma}\bigl(\bP(\bU)\bigr)
	+
	\bS_{\Phi}\bigl(\bP(\bU)\bigr).
\end{equation}
Starting from the current state $\bU^n$, we first calculate the midpoint state $\bU^{n+1/2}$. Then, we use this midpoint state to calculate the new state $\bU^{n+1}$:
\begin{equation}
	\begin{aligned}
		\bU^{n+1/2}
		 & =
		\bU^{n}
		+
		\frac{\Delta t}{2}
		\mathcal{L}(\bU^{n}),
		\\
		\bU^{n+1}
		 & =
		\bU^{n}
		+
		\Delta t\,
		\mathcal{L}(\bU^{n+1/2}).
		\label{eq:RK2}
	\end{aligned}
\end{equation}

When self-gravity is enabled, the potential is recomputed from the
intermediate primitive state before the second Runge--Kutta stage. The mass and spin of the black hole, metric coefficients, and geometrical
source terms are also updated at the intermediate stage.

\subsection{Primitive variable recovery}
\label{subsect:prim_recov}
The primary primitive-variable recovery method is performed by the two-dimensional method \citep{HARM2006}. We define the auxiliary energy--momentum density four-vector
$\mathcal{Q}_{\mu}$, measured by the normal observer, and its spatial
projection $\tilde{\mathcal{Q}}^{\mu}$, given by:

\begin{equation}
	\mathcal{Q}_{\mu} = -n_\nu T^{\nu}_{~\mu},
	\qquad
	\tilde{\mathcal{Q}}^{\mu} = \gamma^\mu_{~\nu} \mathcal{Q}^{\nu}.
\end{equation}
We introduce the auxiliary variable \mbox{$W \equiv \Gamma^2 w$}, where \mbox{$w\equiv \rho + u + p$} is the enthalpy density. Additionally, we define the magnetic field four-vector measured by the normal observer as \mbox{$\mathfrak{B}^\mu = -n_\nu {^*F}^{\mu\nu}$}. The primitive-variable recovery is then reduced to solving the following two equations for
$W$ and $v^2$:
\begin{equation}
	\left\{
	\begin{aligned}
		\tilde{\mathcal{Q}}^2
		 & = v^2 \left( \mathfrak{B}^2 + W \right)^2
		- \frac{
			\left(\mathcal{Q}_\mu \mathfrak{B}^\mu\right)^2
			\left( \mathfrak{B}^2 + 2W \right)
		}{W^2},                                               \\
		\mathcal{Q}_\mu n^\mu
		 & =- \frac{\mathfrak{B}^2}{2} \left( 1 + v^2 \right)
		+ \frac{
			\left(\mathcal{Q}_\mu \mathfrak{B}^\mu\right)^2
		}{2 W^2}
		- W + p(u,\rho).
	\end{aligned}
	\right.
\end{equation}
For the ideal-gas equation of state, the pressure can be expressed as a function of the two variables above:
\begin{equation}
	p =\left(\frac{ \gamma_{\rm ad} -1}{ \gamma_{\rm ad}}\right) \left [ W \left (1-v^2 \right ) - D\sqrt{1-v^2}\right] ,
\end{equation}
where \mbox{$D=\rho \Gamma$}. The system of equations is solved using the Newton--Raphson method. {In addition, we evolve the entropy variable \mbox{$K=p/\rho^{\gamma_{\rm ad}}$} as an auxiliary conserved variable.} If the two-dimensional recovery fails, the code first uses an isentropic one-dimensional root-finding method for $v^2$. If this also fails, it uses a second isentropic one-dimensional method to solve for $W$. If all recovery methods fail, we use neighbouring-cell averaging of hydrodynamic variables, {with recovery flags exchanged between MPI subdomains.} {Other primitive recovery methods have also been proposed in previous studies \citep{Siegel2018,Kastaun2021}.} After recovery, we apply numerical floors:
\begin{equation}
	\begin{aligned}
		\rho_{\rm atm}
		 & =
		\max\left(
		10^{-8} r^{-2},
		10^{-20}
		\right), \\
		u_{\rm atm}
		 & =
		\max\left(
		10^{-10} r^{-2\gamma_{\rm ad}},
		10^{-20}
		\right).
	\end{aligned}
\end{equation}
{These are the default floor values in FUGUE and can be adjusted for different simulations and scientific goals. For example, higher floors are used in the GRMHD code comparison in Sect.~\ref{sec:Porth}}. The final floors also include magnetization limits. In particular, we
limit the magnetization to \mbox{$\sigma=b^2/\rho \leq 50$} and the internal-energy magnetization to \mbox{$b^2/u \leq 2500$}, a limit on the internal-energy-to-density ratio \mbox{$u/\rho \leq 50$}, and the Lorentz factor ceiling \mbox{$\Gamma \leq 50$}. These values are used as the default settings in our simulations. Additionally, in the code, we use drifting floors in highly magnetized regions, where numerical floors are applied. The velocity is adjusted toward the drift-frame velocity, which helps to avoid unphysical velocities after the floors add mass or internal energy.

At the radial boundaries, the code copies primitive variables into ghost zones. The code uses three ghost zones in each spatial direction for PPM reconstruction and two for PLM. The condition for no-inflow is applied to prevent matter from entering the computational domain radially. At the polar boundaries, reflecting boundary conditions are used with the mapping \mbox{$\phi\rightarrow\phi+\pi$} and the correct sign changes for vector components. The azimuthal boundary conditions are periodic. The code solves the equations of general relativistic magnetohydrodynamics (GRMHD) in both axisymmetric 2.5D and fully three-dimensional configurations.
\section{Numerical validation}
\label{sec:validation}
\begin{table*}
	\caption{\label{tab:all_shock_tubes_ic}Initial data for the one-dimensional relativistic shock-tube tests.
		The discontinuity is located at \mbox{$x_0=0.5$} in the domain \mbox{$x\in[0,1]$}.
		The labels S1--S3 denote the hydrodynamic shock tubes, while B1--B5 denote the Balsara magnetohydrodynamic tests. The adiabatic index is chosen as \mbox{$\gamma_{\rm ad}=5/3$} for S1 and S2, \mbox{$\gamma_{\rm ad}=4/3$} for S3, \mbox{$\gamma_{\rm ad}=2$} for B1, and \mbox{$\gamma_{\rm ad}=5/3$} for B2--B5.}
	\centering
	\setlength{\tabcolsep}{2.5pt}
	\renewcommand{\arraystretch}{1.15}

	\begin{tabular*}{\textwidth}{@{\extracolsep{\fill}}|C|CC|CC|CC|CC|CC|CC|CC|CC|}
		\specialrule{0.4pt}{0pt}{2pt}
		\specialrule{0.4pt}{0.3pt}{0pt}
		\noalign{\vskip 0.3mm}

		Test:
		& \multicolumn{2}{c|}{S1}
		& \multicolumn{2}{c|}{S2}
		& \multicolumn{2}{c|}{S3}
		& \multicolumn{2}{c|}{B1}
		& \multicolumn{2}{c|}{B2}
		& \multicolumn{2}{c|}{B3}
		& \multicolumn{2}{c|}{B4}
		& \multicolumn{2}{c|}{B5} \\
		\hline

		& L & R
		& L & R
		& L & R
		& L & R
		& L & R
		& L & R
		& L & R
		& L & R \\
		\hline

		$\rho$
		& $1.0$ & $1.0$
		& $1.0$ & $10.0$
		& $1.0$ & $1.0$
		& $1.0$ & $0.125$
		& $1.0$ & $1.0$
		& $1.0$ & $1.0$
		& $1.0$ & $1.0$
		& $1.08$ & $1.0$ \\
		\hline

		$p$
		& $10^{3}$ & $10^{-2}$
		& $10$ & $20$
		& $1$ & $10$
		& $1.0$ & $0.1$
		& $30.0$ & $1.0$
		& $1000.0$ & $0.1$
		& $0.1$ & $0.1$
		& $0.95$ & $1.0$ \\
		\hline

		$v_x$
		& $0$ & $0$
		& $-0.6$ & $0.5$
		& $0.9$ & $0$
		& $0.0$ & $0.0$
		& $0.0$ & $0.0$
		& $0.0$ & $0.0$
		& $0.999$ & $-0.999$
		& $0.4$ & $-0.45$ \\
		\hline

		$v_y$
		& -- & --
		& -- & --
		& -- & --
		& $0.0$ & $0.0$
		& $0.0$ & $0.0$
		& $0.0$ & $0.0$
		& $0.0$ & $0.0$
		& $0.3$ & $-0.2$ \\
		\hline

		$v_z$
		& -- & --
		& -- & --
		& -- & --
		& $0.0$ & $0.0$
		& $0.0$ & $0.0$
		& $0.0$ & $0.0$
		& $0.0$ & $0.0$
		& $0.2$ & $0.2$ \\
		\hline

		$B^x$
		& -- & --
		& -- & --
		& -- & --
		& $0.5$ & $0.5$
		& $5.0$ & $5.0$
		& $10.0$ & $10.0$
		& $10.0$ & $10.0$
		& $2.0$ & $2.0$ \\
		\hline

		$B^y$
		& -- & --
		& -- & --
		& -- & --
		& $1.0$ & $-1.0$
		& $6.0$ & $0.7$
		& $7.0$ & $0.7$
		& $7.0$ & $-7.0$
		& $0.3$ & $-0.7$ \\
		\hline

		$B^z$
		& -- & --
		& -- & --
		& -- & --
		& $0.0$ & $0.0$
		& $6.0$ & $0.7$
		& $7.0$ & $0.7$
		& $7.0$ & $-7.0$
		& $0.3$ & $0.5$ \\

		\specialrule{0.4pt}{0pt}{0pt}
	\end{tabular*}
\end{table*}
We present below a set of numerical tests of the code, performed directly on the GPU version. We begin with special relativistic (SR) hydrodynamics tests, followed by SRMHD tests. Then, we test GRMHD, and finally, we validate the weak-field multipole self-gravity implementation.
\subsection{Relativistic hydrodynamic shock-tube tests}
\begin{figure*}
	\centering
	\includegraphics[width=1\linewidth]{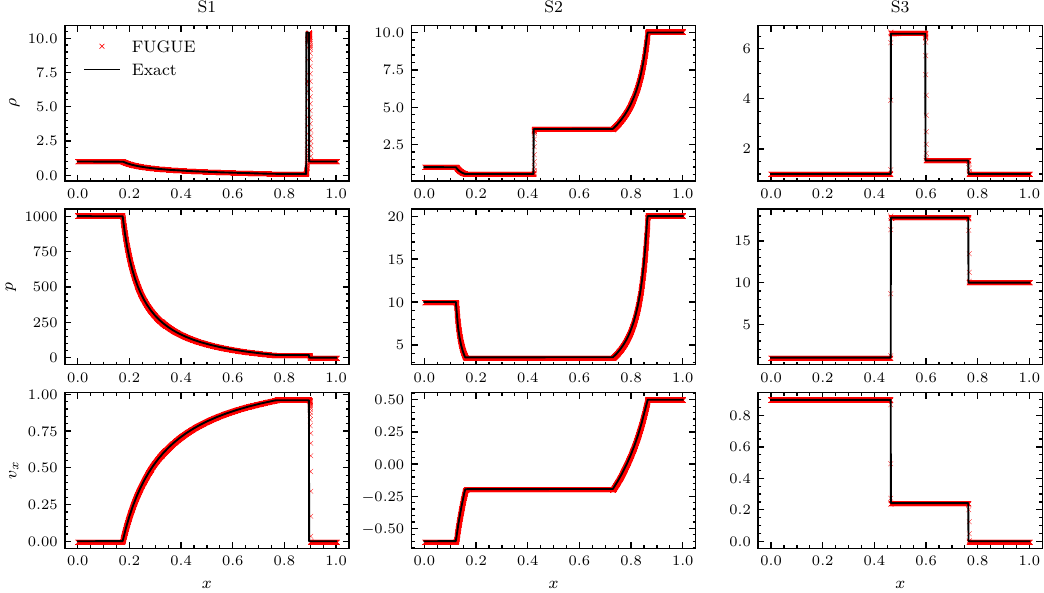}
	\caption{Relativistic hydrodynamic shock-tube tests. The figure compares the {FUGUE} results with the exact solutions for the \mbox{$\mathrm{S1-S3}$} problems. The calculations use PPM reconstruction, the resolution of \mbox{$N=8192$}, the Courant factor \mbox{${\rm CFL}=0.4$}, and the final time of \mbox{$t=0.4$}. The code accurately captures shocks and contact discontinuities in density, pressure, and velocity.} {}
	\label{fig:THOMSON}
\end{figure*}
\label{ST_REL}
We begin testing our code with one-dimensional special-relativistic shock-tube tests introduced by Thompson \citep{Thompson_1986}. We adopt the Minkowski metric in Cartesian coordinates. The discontinuity is located in the middle of the grid at \mbox{$x=0.5$}. The initial data are presented in Table~\ref{tab:all_shock_tubes_ic}. For all tests (\mbox{$\mathrm{S1-S3}$}), we use the resolution of \mbox{$N=8192$} cells. We use outflow boundary conditions at both ends of the domain. The chosen Riemann solver is HLLE. For the reconstruction, we use PPM with shock flattening. The Courant factor is set to \mbox{${\rm CFL}=0.4$}. The exact results are obtained using the exact Riemann solver \citep{Rezzolla2003}. The final time of the evolution is set to \mbox{$t=0.4$} for all shock-tubes. In Fig.~\ref{fig:THOMSON} are shown the results obtained with FUGUE, and the exact solutions for tests \mbox{$(\mathrm{S1-S3})$}. The code accurately reproduces density, pressure, and velocity profiles, including shocks and contact discontinuities, without spurious oscillations.
\subsection{Kelvin--Helmholtz instability}
\label{KH}
\begin{figure*}
	\centering
	\includegraphics[width=0.995\linewidth,trim=0 0 0 0,clip]{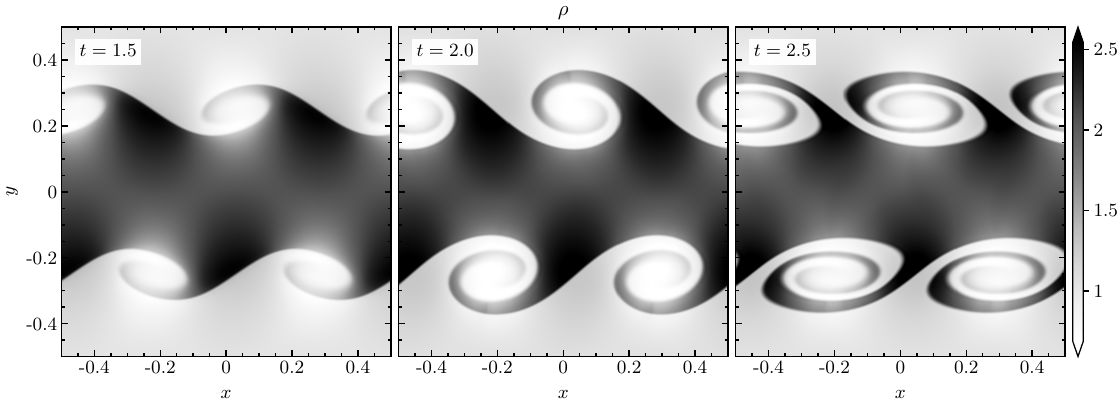}
	\caption{Kelvin--Helmholtz instability test. The panels show density maps at \mbox{$t=1.5,\,2.0,\,2.5$}. The resolution is set to \mbox{$1024\times1024$}. The reconstruction scheme is chosen as PLM with the \mbox{\textsc{MC}} flux limiter. The Courant factor is set to \mbox{${\rm CFL}=0.8$}. The code reproduces the characteristics of rolling up vortices and their evolution in time.}
	\label{fig:KH}
\end{figure*}

We consider the two-dimensional Kelvin--Helmholtz instability in Cartesian coordinates in Minkowski spacetime. We test whether the code can capture and reproduce the characteristic Kelvin--Helmholtz vortices. The initial setup is based on a similar test described by \citet{AsterX}, with a divergence-free velocity perturbation. The computational domain is \mbox{$x,y\in[-0.5,0.5]$}. We use a uniform
\mbox{$1024\times1024$} grid with periodic boundary conditions. The adiabatic index is \mbox{$\gamma_{\rm ad} = 5/3$}, and the gas pressure is constant with the value \mbox{$p = 2.5$}. The magnetic field is set to zero. The central layer, \mbox{$|y|<y_s$}, where \mbox{$y_s=0.25$}, has density \mbox{$\rho=2$} and initial velocity in the $x$-direction \mbox{$v^x_0=0.5$}, while the outer layers have density \mbox{$\rho=1$} and initial velocity in the $x$-direction \mbox{$v^x_0=-0.5$}. This produces two oppositely oriented shear interfaces at \mbox{$y=\pm \,0.25$}. We perturb the flow velocity according to the following prescription:
\begin{equation}
	\begin{split}
		\psi
		=
		\frac{w_0}{k}\cos(kx)
		\biggl[
		 & \exp\left(-\frac{(y-y_s)^2}{2\sigma^2}\right)
		\\
		 & +
			\exp\left(-\frac{(y+y_s)^2}{2\sigma^2}\right)
			\biggr].
	\end{split}
\end{equation}
This choice makes the initial perturbation divergence-free. We use \mbox{$k=4\pi$}, \mbox{$\sigma=0.05/\sqrt{2}$}, and
\mbox{$w_0=0.015$}. The initial velocity is then
\mbox{$v^x=v^x_0+\delta v^x$} and \mbox{$v^y=\delta v^y$}. The velocity perturbations are defined by \mbox{$\delta v^x=\partial_y\psi$} and \mbox{$\delta v^y=-\partial_x\psi$}, which ensures \mbox{$\partial_x\delta v^x+\partial_y\delta v^y=0$}. The system is evolved until \mbox{$t=2.5$}. The Riemann solver is set to HLLE, with the PLM reconstruction and the \mbox{\textsc{MC}} flux limiter. The Courant factor is set to \mbox{${\rm CFL}=0.8$}. In Fig.~\ref{fig:KH} we show the density maps for \mbox{$t=1.5,\, 2.0,\,2.5$}. The density plot shows the two shear layers rolling up into vortices, demonstrating that the code can properly capture the growth of the Kelvin--Helmholtz instability over time. This indicates that numerical diffusion is low enough to resolve instabilities.
\subsection{Relativistic magnetohydrodynamic shock-tube tests}
\label{ST_MAG}
\begin{figure*}
	\centering
	\includegraphics[width=1\linewidth]{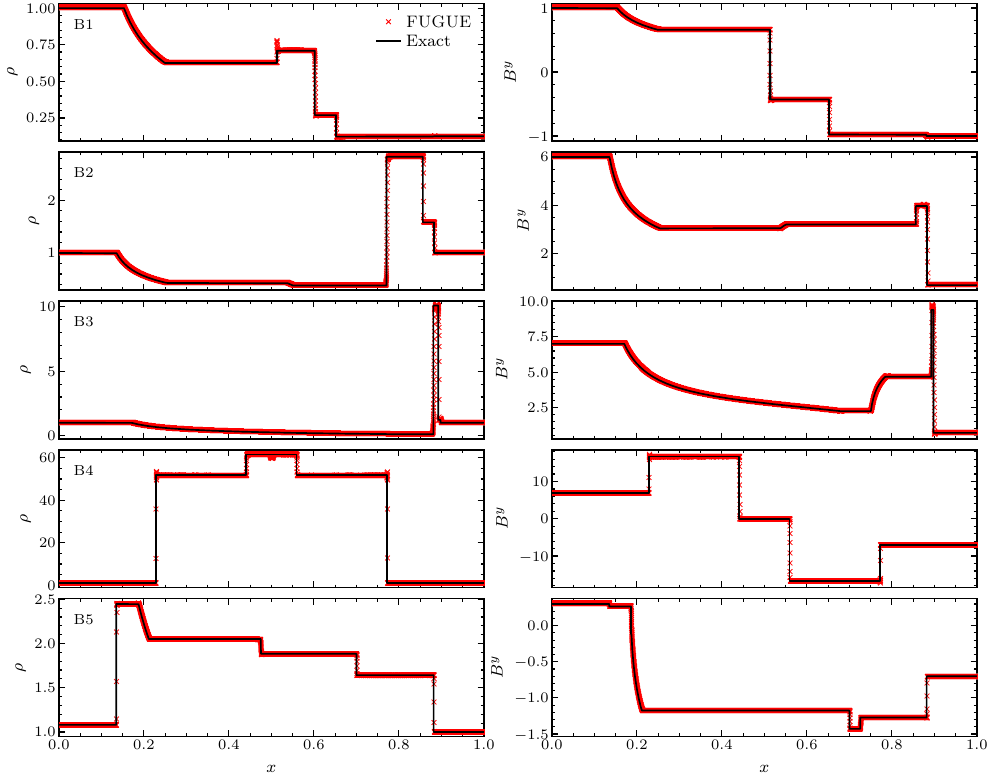}
	\caption{
		Relativistic magnetohydrodynamic Balsara shock-tube tests. Red crosses show the numerical solutions obtained with {FUGUE}, whereas the black curves show the corresponding exact solutions. The left panels show the density $\rho$, and the right panels show the transverse magnetic field component $B^y$.
	}
	\label{fig:balsara}
\end{figure*}

We start testing the implementation of relativistic magnetohydrodynamics using Balsara shock-tubes \citep{Balsara}. The initial conditions are presented in Table~\ref{tab:all_shock_tubes_ic}. We use Minkowski spacetime in Cartesian coordinates. We use the PLM reconstruction method, with the \mbox{\textsc{MC}} flux limiter. The Riemann solver is chosen to be Lax--Friedrichs. The grid resolution is set to \mbox{$N=8192$}, with outflow boundary conditions. The Courant factor is set to \mbox{${\rm CFL}=0.4$}. The five Balsara problems have different levels of difficulty. For instance, the shock-tube B3 has a very large pressure jump, from \mbox{$p=1000$} to \mbox{$p=0.1$}, together with a strong magnetic field, whereas the shock-tube B4 is the most relativistic case, with two flows moving toward each other at \mbox{$v^x=\pm0.999$}, and a transverse magnetic field that changes sign across the initial discontinuity.. Tests B1--B4 are evolved to \mbox{$t=0.4$}, while test B5 is evolved to \mbox{$t=0.55$}. In Fig.~\ref{fig:balsara}, we present profiles of density $\rho$ and the magnetic field $B^y$. The numerical results are compared with the exact solutions \citep{Giacomazzo2006}. These tests include strong shocks and sharp changes in the fluid variables. They test whether the numerical methods work correctly in difficult cases. Even in the most extreme cases, the code provides solutions in close agreement with the exact solutions. We do not observe overshooting or oscillations near discontinuities.
\subsection{Linear waves}
\begin{figure}
	\centering
	\includegraphics[width=0.99\linewidth]{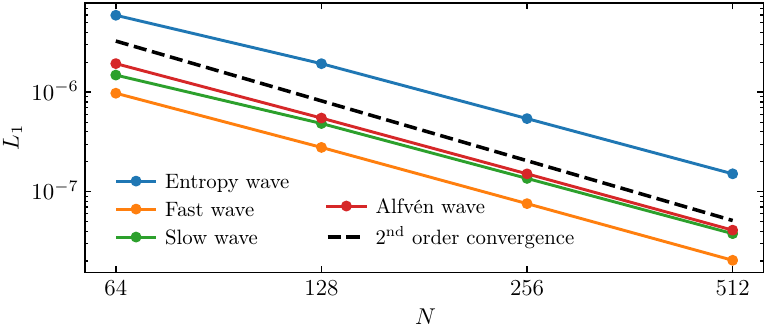}
	\caption{
		The absolute $L_1$ error for entropy, fast magnetosonic, slow magnetosonic, and Alfvén waves. Each wave is initialized with a small-amplitude perturbation from the corresponding eigenvector. The error decreases at approximately second order with increasing resolution.}
	\label{fig:linear}
\end{figure}
\label{linear}
We test the convergence order of the code using standard one-dimensional linear-wave problems in the Cartesian Minkowski spacetime. We consider the entropy, fast magnetosonic, slow magnetosonic, and Alfvén waves \citep{Stone2008,H-AMR}. The computational grid is set to \mbox{$x\in[-0.5,0.5]$} with periodic boundary conditions. We test resolutions \mbox{$N\in\{64,\,128,\,256,\,512\}$}. We initialize each type of wave by adding a small-amplitude perturbation proportional to the eigenvector of the corresponding linear mode. The amplitude is chosen to be \mbox{$A=10^{-4}$}. We use the PLM method for reconstruction with the \mbox{\textsc{MINMOD}} flux limiter. The Riemann solver is HLLE. The Courant factor is set to \mbox{${\rm CFL}=0.4$}. In Fig.~\ref{fig:linear}, we show the absolute $L_1$ error after one period. The error is the mean absolute difference between the final and initial states, averaged over all grid cells and all primitive variables with non-zero perturbations. The test shows second-order convergence for all tested waves.
\subsection{Blast-wave test}
\label{blast}
\begin{figure*}
	\centering
	\includegraphics[width=0.99\linewidth]{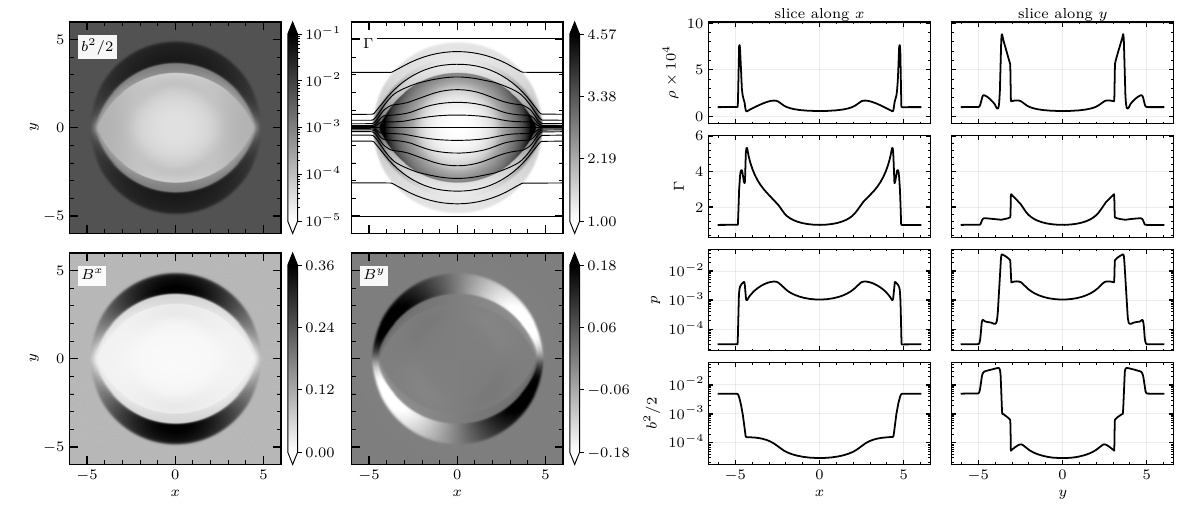}
	\caption{Two-dimensional cylindrical relativistic magnetohydrodynamics blast-wave test. The left panels show maps of magnetic pressure $b^2/2$, Lorentz factor $\Gamma$, and magnetic field components $B^x$ and $B^y$ at the final time \mbox{$t=4.0$}. The right panels show one-dimensional slices along the $x$ and $y$ directions. Black lines in the Lorentz-factor panel show the magnetic field.}
	\label{fig:blast_wave_combined}
\end{figure*}

We consider the two-dimensional cylindrical blast wave in a uniformly magnetized medium \citep{Komissarov1999,Zanna2003}. The initial conditions describe a dense and over-pressured circular region surrounded by low-density ambient medium. In contrast to the one-dimensional shock-tube tests, there is no analytic solution; therefore, the comparison is usually based on reproducing the characteristic solution structures reported in the literature. The computational domain covers \mbox{$x,y\in[-6,6]$}, with the resolution set to \mbox{$1024\times1024$} cells. The adiabatic index is set to \mbox{$\gamma_{\rm ad}=4/3$}, the final time is \mbox{$t=4.0$}, and the Courant factor is
\mbox{${\rm CFL}=0.4$}. As the reconstruction method, we use PLM with the \mbox{\textsc{MINMOD}} flux limiter. The approximate Riemann solver is chosen to be HLLE. Initially, the fluid
velocity is set to zero everywhere, while the magnetic field is uniform and
given by \mbox{$B^x=0.1$}, \mbox{$B^y=0$}, and \mbox{$B^z=0$}. The cylinder is specified by the smoothing radii
\mbox{$r_\mathrm{in}=0.8$} and \mbox{$r_\mathrm{out}=1.0$}. Inside the cylinder, we set the density
\mbox{$\rho_\mathrm{in}=10^{-2}$} and the pressure \mbox{$p_\mathrm{in}=1.0$}, while in the outer
medium we set \mbox{$\rho_\mathrm{out}=10^{-4}$} and
\mbox{$p_\mathrm{out}=3\times10^{-5}$}. The density is set in the zone \mbox{$r_\mathrm{in}\leq r\leq r_\mathrm{out}$} by the following equation:
\begin{equation}
	\rho(r)
	=
	\exp\left[
		\frac{
			(r_{\mathrm{out}} - r)\ln\rho_{\mathrm{in}}
			+
			(r - r_{\mathrm{in}})\ln\rho_{\mathrm{out}}
		}{
			r_{\mathrm{out}} - r_{\mathrm{in}}
		}
		\right].
\end{equation}
In Fig.~\ref{fig:blast_wave_combined}, we show the state at the final time
\mbox{$t=4.0$}. The left panels show two-dimensional maps of the magnetic pressure
$b^2/2$, Lorentz factor $\Gamma$, and magnetic field components $B^x$ and $B^y$.
The right panels show one-dimensional slices along the $x$ and $y$ axes with the density $\rho$. The code reproduces results known from the literature, and remains numerically stable.
\subsection{Magnetic rotator}
\label{magrot}
\begin{figure*}
	\centering
	\includegraphics[width=0.99\linewidth]{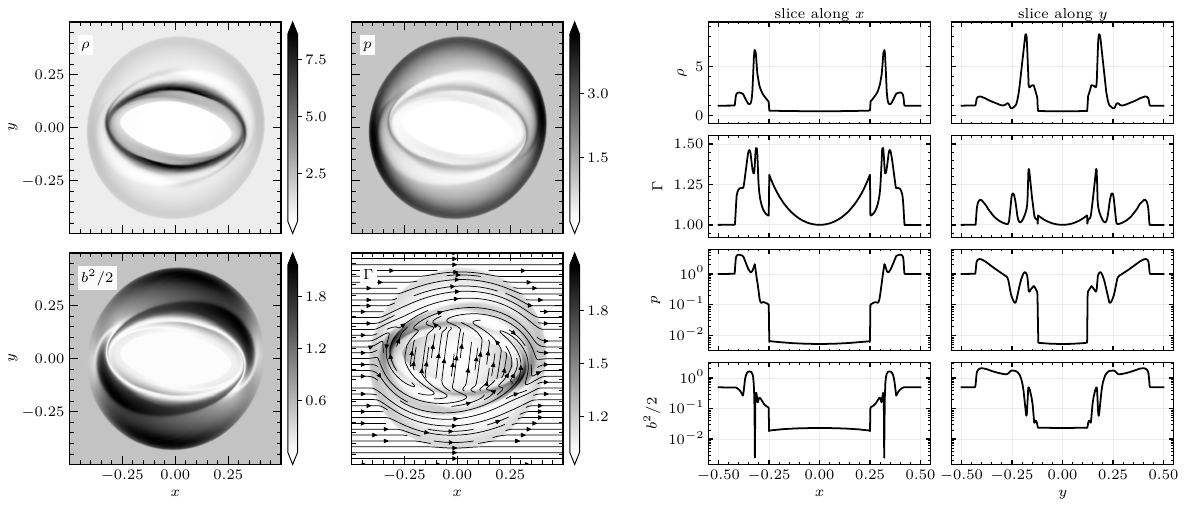}
	\caption{
		Relativistic magnetic-rotor test. The left panels show two-dimensional maps of density $\rho$, pressure $p$, magnetic pressure $b^2/2$, and Lorentz factor $\Gamma$ at the final time. The right panels show one-dimensional slices along the $x$ and $y$ directions. The result shows that the code can handle strong magnetic stresses and remains stable.}
	\label{fig:magrot}
\end{figure*}

We consider the relativistic magnetic rotor test \citep{BALSARA1999270,Toth2000}. The test describes a dense rotating cylinder in a uniform magnetic
field. The computational domain is \mbox{$x,y\in[-0.5,0.5]$}. We use the resolution of \mbox{$1024\times1024$} cells. The ideal-gas equation of state with \mbox{$\gamma_{\rm ad}=5/3$} is adopted. The final time is \mbox{$t=0.4$}, and the Courant factor is \mbox{${\rm CFL}=0.25$}. We use the HLLE Riemann solver with PLM reconstruction, with the \mbox{\textsc{MINMOD}} flux limiter. The rotor has a radius of \mbox{$R=0.1$}. Inside the rotor, the density is \mbox{$\rho_\mathrm{in}=10$}, while outside it is set to
\mbox{$\rho_\mathrm{out}=1$}. The pressure is uniform and equal to \mbox{$p=1$}. The magnetic field is also uniform and is initially given by
\mbox{$B^x=1$}, \mbox{$B^y=B^z=0$}. Inside the rotor, the gas rotates with angular velocity \mbox{$\Omega=9.95$}, so the velocities are given by \mbox{$v^x=-\Omega y$} and \mbox{$v^y=\Omega x$}. This gives a maximum speed close to \mbox{$v_\mathrm{max}=0.995$} at the edge of the rotor. The fluid outside the rotor is at
rest. In Fig.~\ref{fig:magrot}, we show the result at the final time. The maps show the density, pressure, magnetic pressure $b^2/2$, and Lorentz factor $\Gamma$, with the magnetic field lines. On the right side, the slices are shown along $x$ and $y$. The results of magnetic rotator test show the structures known from the literature, indicating that the code can handle extreme rotating magnetized fluid and stays stable.

\subsection{Advection loop test}
\label{advection}
\begin{figure*}
	\centering
	\includegraphics[width=0.99\linewidth]{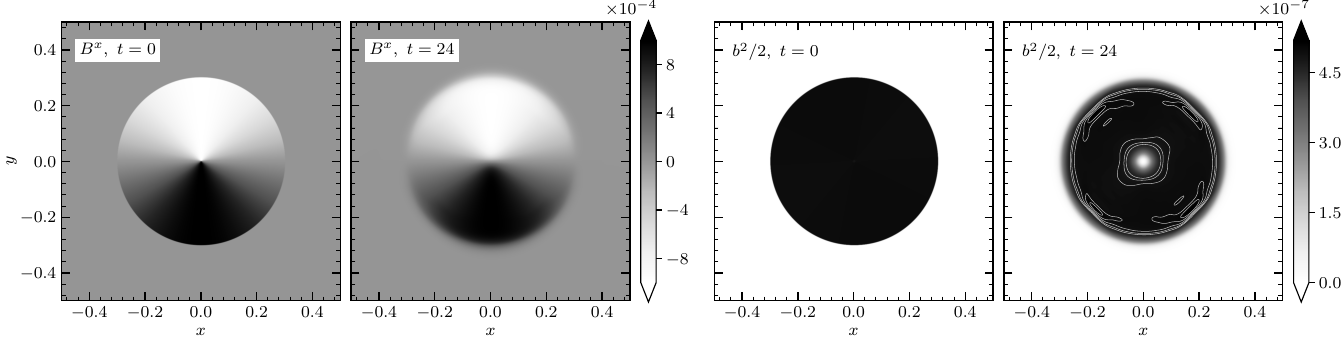}
	\caption{Magnetic-loop advection test. The figure shows the magnetic field component $B^x$ and magnetic pressure $b^2/2$ at the initial and final times. After two box crossings in the $x$ direction and one in the $y$ direction, the loop remains close to its initial shape. This demonstrates that the code advects weak magnetic fields correctly with only small numerical diffusion.}
	\label{fig:advloop}
\end{figure*}

We consider the magnetic-loop advection test \citep{Gardiner2005} in Cartesian coordinates in Minkowski spacetime. The test describes the advection of a weak magnetic loop by a uniform fluid flow. The computational domain is \mbox{$x,y\in[-0.5,0.5]$}. We use periodic boundary conditions in all directions. The grid resolution is \mbox{$1024\times1024$} cells. The gas is initially uniform, with \mbox{$\rho=1$}, \mbox{$p=3$}, and adiabatic index \mbox{$\gamma_{\rm ad}=5/3$}. The velocity is constant and is set to \mbox{$v^x=1/12$}, \mbox{$v^y=1/24$}, and \mbox{$v^z=1/24$}. We use the HLLE Riemann solver, with PLM reconstruction and the \mbox{\textsc{MINMOD}} flux limiter. The CFL is set to \mbox{${\rm CFL}=0.4$}. We set \mbox{$A_z=A_\mathrm{loop}\max(R_\mathrm{loop}-r,0)$}, where \mbox{$r=\sqrt{x^2+y^2}$}, \mbox{$A_\mathrm{loop}=10^{-3}$}, and
\mbox{$R_\mathrm{loop}=0.3$}. The final time is \mbox{$t=24$}, which corresponds to two box crossings in the $x$ direction and one box crossing in the $y$ direction. In Fig.~\ref{fig:advloop}, we show $B^x$ and the magnetic pressure $b^2/2$ at the initial and final times. This test shows that the code transports weak magnetic fields correctly
and keeps the magnetic structure stable over time, with only small numerical diffusion.
\subsection{Magnetized Bondi--Michel accretion}
\label{bondi}
\begin{figure*}
	\centering
	\includegraphics[width=0.99\linewidth]{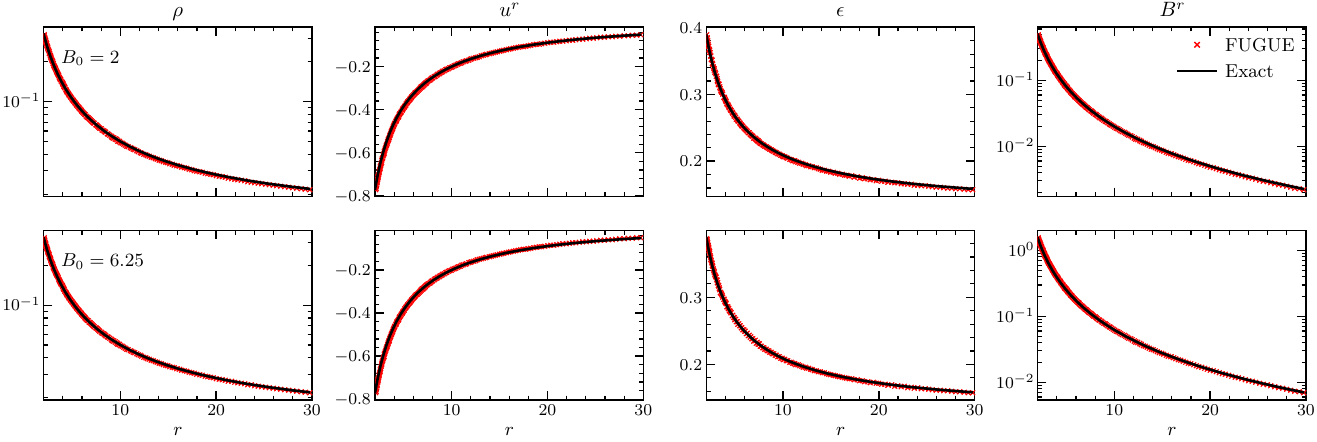}
	\caption{Magnetized Bondi--Michel accretion test in Schwarzschild
		spacetime for two magnetic field strengths, \mbox{$B_0=2$} and
		\mbox{$B_0=6.25$}. The shown variables are the rest-mass density $\rho$, the radial velocity  $u^r$, the specific internal energy $\epsilon$, and the magnetic field $B^r$. The figure shows radial profiles from the simulation and compares them with the exact solutions. The numerical solutions follow the analytic profile, which shows that the code can reproduce a steady spherical accretion flow.}
	\label{fig:magbondi}
\end{figure*}

After validating special relativistic magnetohydrodynamics, we proceed to test general relativistic magnetohydrodynamics (GRMHD). We start with the magnetized Bondi--Michel accretion
solution in Schwarzschild spacetime \citep{Michel1972,Fedrigo2025}. The problem has a well-known stationary solution; therefore, it is a useful test for code validation. The black hole mass is set to \mbox{$M_{\mathrm{BH}}=1$}, and the spin is set to \mbox{$a_{\star}=0$}. We use an ideal-gas equation of state with \mbox{$\gamma_{\rm ad}=4/3$}. The Courant factor is set to \mbox{${\rm CFL}=0.4$}. We adopt spherical coordinates and use a resolution of \mbox{$(N_r\times N_\theta\times N_\phi)=(256\times128\times64)$}. The Riemann solver is HLLE with PPM as the reconstruction method. The outer boundary is set to $r_{\mathrm{out}}=100$. At this boundary, we use the Bondi solution for the inflow.
The sonic radius is chosen to be \mbox{$r_s=8$}, the mass accretion rate is \mbox{$\dot{M}=-1$}, and the density at the sonic radius is \mbox{$\rho(r_s)=1/16$}. The hydrodynamic solution is obtained by solving the mass-conservation equation above and below the sonic radius, given by:
\begin{equation}
	\rho(r) =\frac{\dot{M}}{r^2 u^r},
	\label{eq:masscon}
\end{equation}
and by the relativistic Bernoulli equation \citep{Michel1972}, describing conservation of the relativistic energy flux defined as follows:
\begin{equation}
	\dot{E}=r^2 u^r\left(\rho + \frac{\gamma_{\rm ad}\, p}{\gamma_{\rm ad}-1}\right)\left[1-\frac{2M}{r} +(u^r)^2\right]^{1/2},
\end{equation}
assuming that $\dot{E}$ is constant. Here, $\dot{M}$ and $\dot{E}$ are signed fluxes per unit solid angle. At each radius, $u^r$ is found as the root of the Bernoulli equation using the secant method, and the density from mass conservation. The internal energy is then calculated from the polytropic relation. On top of the Bondi solution, we add a purely radial magnetic field: \mbox{$B^r = {B_0}/{r^2}$}, with \mbox{$B^\theta = B^\phi = 0$}. We perform the test for two magnetic field strengths \mbox{$B_0 = 2$} and \mbox{$B_0 = 6.25$}. The system is evolved numerically and compared with the initial analytic profiles. We evolve the system to \mbox{$t=100$}. In Fig.~\ref{fig:magbondi}, the radial profiles of the density $\rho$, radial velocity $u^r$, specific internal energy \mbox{$\epsilon = u/\rho$}, and magnetic field $B^r$ are shown. The code preserves the stationary solution for two values of the magnetic field strengths. This test checks both the HD and MHD parts of the GRMHD solver.
\subsection{Fishbone--Moncrief torus}
\label{subsec:FM}
\begin{figure}
	\centering
	\includegraphics[width=0.99\linewidth]{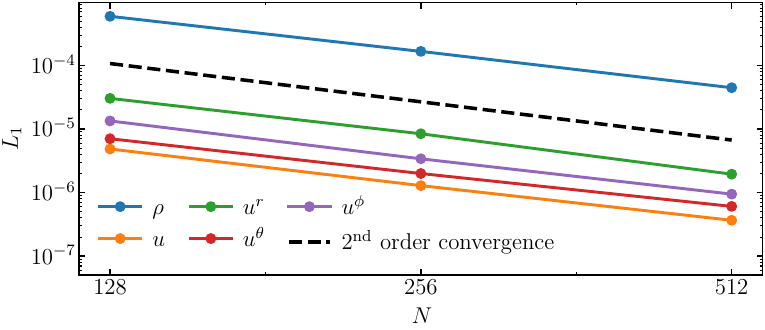}
	\caption{Convergence of the Fishbone--Moncrief torus test. The figure shows
		the absolute $L_1$ errors of the primitive variables: the rest-mass density, internal energy density, and the spatial components of the fluid four-velocity, as a function of the grid
		resolution \mbox{$N\times N$}. The black dashed line indicates second-order convergence.}
	\label{fig:FM}
\end{figure}

We next test the GR hydrodynamics implementation in the Kerr spacetime. We use the Fishbone--Moncrief torus \citep{Fishbone1976,HARM2003}, which is in hydrostatic equilibrium (in the absence of a magnetic field) in the Kerr metric. We test the stability of the torus. The spin of the black hole is chosen as \mbox{$a_{\star}=0.9$}. The inner edge of the torus is at \mbox{$r_{\rm in}=6$}, and the pressure maximum is at \mbox{$r_{\rm max}=12$}. The test is performed in modified Kerr--Schild coordinates with \mbox{$h=0.3$}. We use an ideal-gas equation of state with the adiabatic index \mbox{$\gamma_{\rm ad}=4/3$}. The torus is evolved until \mbox{$t=100$}. For the test, we use the HLLE Riemann solver and PLM reconstruction with MINMOD flux limiter. In Fig.~\ref{fig:FM} we show the absolute $L_1$ errors separately for the rest-mass density, internal energy density, and the spatial components of the fluid
four-velocity, as a function of the grid resolution \mbox{$N\in\{128,\,256,\,512\}$}. We test the torus in axisymmetry with the resolution \mbox{$N_r\times N_\theta$} being the same in radial and polar directions. The errors decrease with approximately second-order accuracy for all variables, indicating that the implementation of GRHD in the Kerr metric, including source terms and geometry, converges to the equilibrium solution.
\subsection{{GRMHD code comparison benchmark}}
\label{sec:Porth}
\begin{figure}
	\centering
	\includegraphics[width=0.99\linewidth]{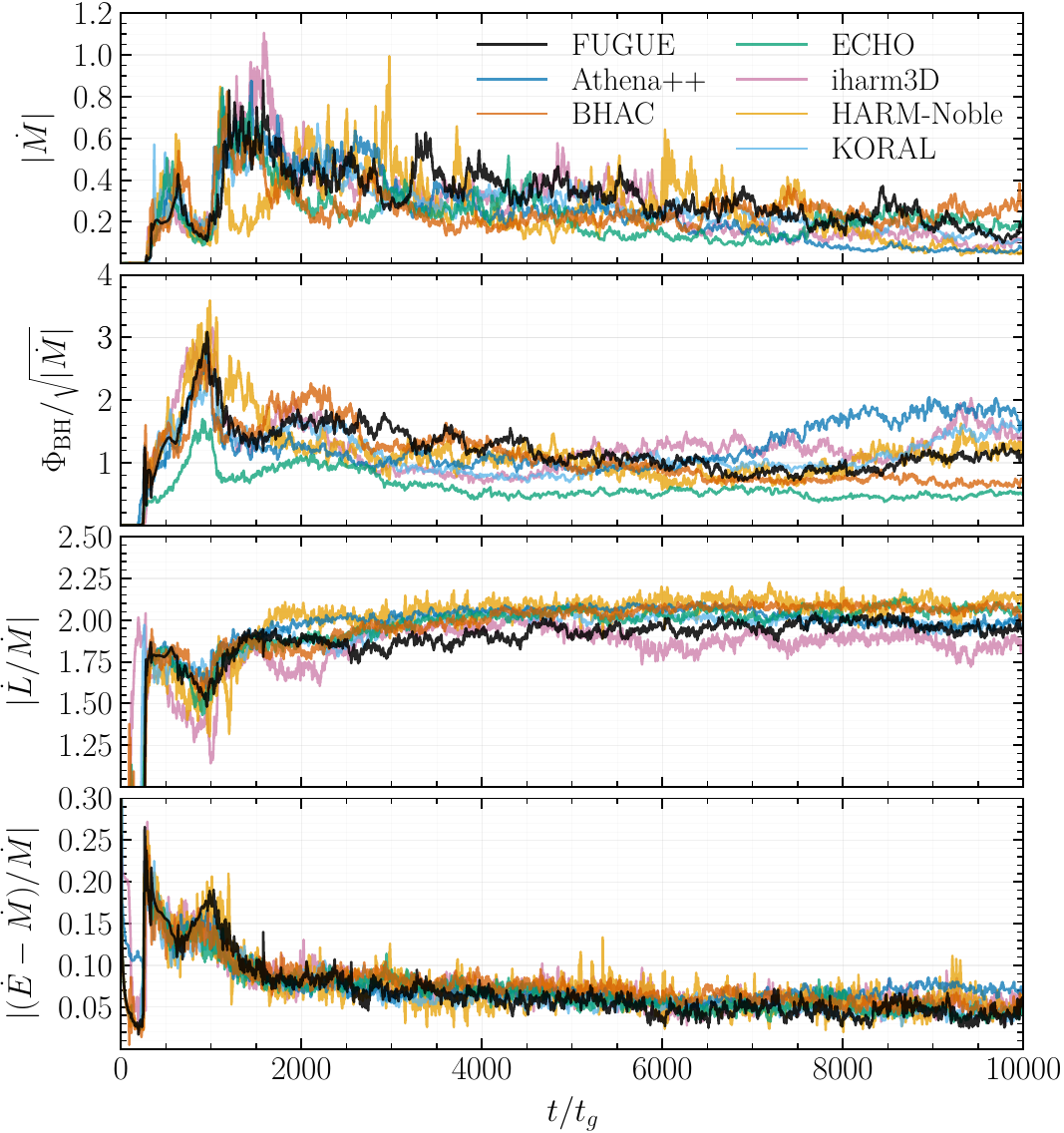}
	\caption{{Comparison of {FUGUE} with the other GRMHD code results at a resolution of $192^3$. The panels show the horizon diagnostics: the mass accretion rate, the dimensionless magnetic flux, and the specific angular momentum flux, and the reduced energy flux.}}
	\label{fig:Porth_comparison}
\end{figure}

{The comparison between various GRMHD codes showed that their results are broadly consistent \citep{Porth2019}. We test {FUGUE}, comparing it with state-of-the-art codes. We use the Fishbone--Moncrief initial conditions \citep{Fishbone1976}, with the parameters of the dimensionless black hole spin \mbox{$a_{\ast}=0.9375$}. The torus has an inner radius \mbox{$r_{\rm in}=6\,r_g$} and a pressure maximum at \mbox{$r_{\rm max}=12\,r_g$}. We use \mbox{$\gamma_{\rm ad}=4/3$} and set the initial plasma beta to \mbox{$\beta=100$}. We add a small $2\%$ perturbation to the internal energy. The grid resolution is $192^3$, with the grid extends radially \mbox{$r\in[1.15,50]\,r_g$} and \mbox{$h=1$}. Near the polar axis, we use cylindrified MKS coordinates described in Sect.~\ref{sect:perf}. We use the local Lax--Friedrichs flux and PLM reconstruction with the MC limiter, and evolve the model up to \mbox{$t=10^4\,t_g$}. We use floors of \mbox{$\rho > 10^{-5}r^{-2}$}
and \mbox{$u > 10^{-7}r^{-2\gamma_{\rm ad}}$, the same values as iharm3D in} \citet{Porth2019}. Self-gravity and black hole evolution are turned off. In Fig.~\ref{fig:Porth_comparison}, we present comparison of {FUGUE} with other codes, comparing horizon diagnostics, namely the mass accretion rate (Eq.~\ref{eq:dotM}), the dimensionless magnetic flux (Eqs.~\ref{eq:dotM} and \ref{eq:phiBH}), the specific angular momentum flux (Eq.~\ref{eq:spin}), and the reduced energy flux (see Eq.~\ref{eq:edot}). Apart from the fluctuations expected from MRI-driven turbulence, the evolution proceeds similarly to that of other numerical codes; we do not observe large deviations from the behavior of other codes: Athena++ \citep{White2016}, BHAC \citep{Porth2017}, ECHO \citep{Zanna2007}, HARM-Noble \citep{HARM2003,HARM2006, Noble2009}, iharm3D \citep{HARM2003,HARM2006, Noble2009,Prather2021}, and KORAL \citep{Sadowski2013,Sadowski2014}. We note that this is not a strict one-to-one comparison, as the codes use different numerical schemes, reconstruction methods, and grids. In Table~\ref{tab:fugue_diagnostics}, we present results from FUGUE that are analogous to those in Table~2 of \citet{Porth2019}. Throughout the evolution, entropy fallback was used in \mbox{$\sim0.13\%$} of cells, and not a single cell required the averaging procedure discussed in Sect.~\ref{subsect:prim_recov}.}
\begin{table}
	\centering
	\caption{{Horizon diagnostics for the {FUGUE} simulation at $192^3$ resolution. Maximum values are taken over the full run, while $\langle\cdot\rangle$ denotes averages over \mbox{$5000 \le t/t_g \le 10000$} with standard deviations.}}
	\label{tab:fugue_diagnostics}
	\setlength{\tabcolsep}{5pt}
	\renewcommand{\arraystretch}{1.15}
	\begin{tabular}{lcc}
		\hline\hline
		Diagnostic & $\max$          & $\langle\cdot\rangle$ \\
		\hline
		$|\dot M|$
		           & $0.883$
		           & $0.244\pm0.068$                         \\

		$\Phi_{\rm BH}/\sqrt{|\dot M|}$
		           & $3.087$
		           & $0.989\pm0.129$                         \\

		$|\dot L/\dot M|$
		           & $2.044$
		           & $1.958\pm0.033$                         \\

		$|(\dot E-\dot M)/\dot M|$
		           & $0.331$
		           & $0.047\pm0.009$                         \\
		\hline\hline
	\end{tabular}
\end{table}
\subsection{Monopole and multipole self-gravity tests}
\label{App:monopole}
\begin{figure}
	\centering
	\includegraphics[width=0.99\linewidth]{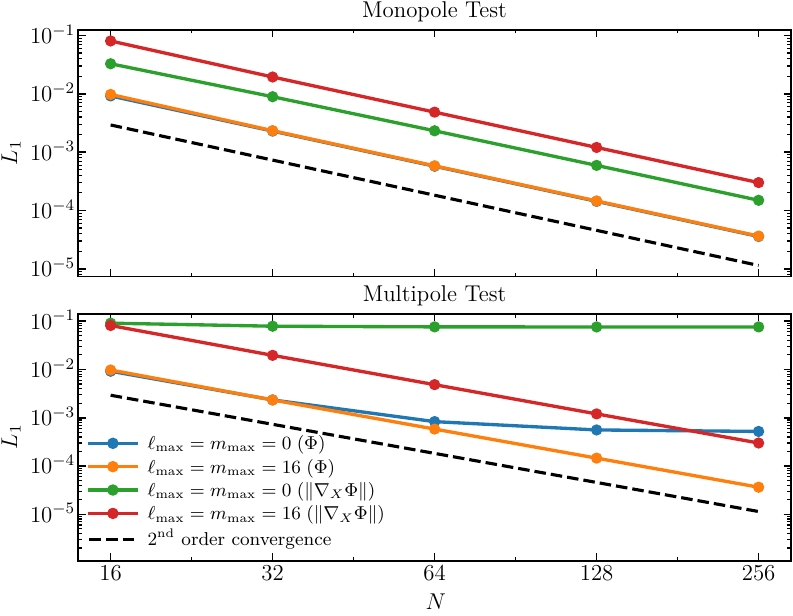}
	\caption{
		Convergence tests of the self-gravity Poisson solver, showing the relative
		$L_1$ errors of the gravitational potential $\Phi$ and its gradient
		\mbox{$\|\nabla_X\Phi\|$}. In the top panel, both the monopole and multipole configurations show second-order convergence for the monopole test. In the bottom panel, the multipole solver converges for the multipole test, whereas the monopole solver cannot fully reproduce the non-spherical field. Dashed lines indicate second-order convergence.
	}
	\label{fig:monopole_test}
\end{figure}

After testing the
GRMHD
solver, we proceed to test the implementation of the self-gravity module. In the code, the multipole Poisson solver operates in modified Kerr--Schild coordinates \citep{McKinney2004,HARM2006}, which introduce a logarithmic radial grid and the parameter $h$ responsible for a denser grid in the equatorial region (Eq.~\ref{eq:MKS}).
In the following tests, we validate our solver using \mbox{$h=0.3$}, creating more difficult conditions for the solver than a uniform grid. We test convergence for \mbox{$N\in\{16,\,32,\,64,\,128,\,256\}$}, keeping three-dimensional resolution as \mbox{$N^3$}. The density distribution, given by an analytical profile, should give a pure monopole (\mbox{$\ell=0,\,m=0$}) potential. The density profile is as follows:
\begin{equation}
	\rho(r) =\rho_0\left({r}/{r_0}\right)^{-q},
\end{equation}
where \mbox{$q=1$}, \mbox{$r_0 =10$}, and \mbox{$\rho_0=1$}. We compare the analytical potential $\Phi$ and its gradient, calculated using Eq.~\ref{eq:grad} with the reconstructed potential obtained using the Poisson solver. We calculate the relative $L_1$ errors of these values. We use two configurations of our solver: a purely monopole solver (\mbox{$\ell_{\mathrm{max}}=m_{\mathrm{max}}=0$}), and the multipole configuration (\mbox{$\ell_{\mathrm{max}}=m_{\mathrm{max}}=16$}). In Fig.~\ref{fig:monopole_test}, the convergence test of the monopole is shown. For a purely monopole gravitational field, we observe second-order convergence for both setups. The higher-order multipoles used in this test can reconstruct the gravitational monopole almost as accurately as the monopole solver; however the potential gradient is slightly more accurate with the monopole configuration.

The main motivation for using the multipole Poisson solver is to go beyond the monopole treatment of self-gravity used in our previous studies \citep{Janiuk2023,Plonka2026}. The monopole solver includes only the spherically symmetric part of the gravitational potential. It cannot describe non-spherical structures, such as spiral density waves or thin disks.

As a second test here, we use a density distribution similar to that used in the monopole test; however, it is multiplied by spherical harmonics, which introduces dependence on the azimuthal and polar angles.
We use the following distribution:
\begin{equation}
	\rho(r,\theta,\phi)
	=
	\rho_0
	(r/r_0)^{-q}
	\left[
		1
		+
		\operatorname{Re}
		\left\{
		Y_{84}(\theta,\phi)
		\right\}
		\right],
	\label{eq:multipole_density}
\end{equation}
where $Y_{84}$ is the spherical harmonic with \mbox{$\ell=8$} and \mbox{$m=4$}. Then, we compare the analytical solution with the numerical %
result obtained using the multipole solver. %
The bottom panel of Fig.~\ref{fig:monopole_test} shows the convergence test. First, we observe the expected behaviour of the monopole solver in the multipole scenario: expected lack of convergence, with the error curve becoming flat. It shows that the solver is unable to reproduce higher-order multipoles. However, with the multipole solver, we obtain convergence close to the second order as the resolution increases.
\subsection{Self-gravitating star}
\begin{figure}
	\centering
	\includegraphics[width=0.99\linewidth]{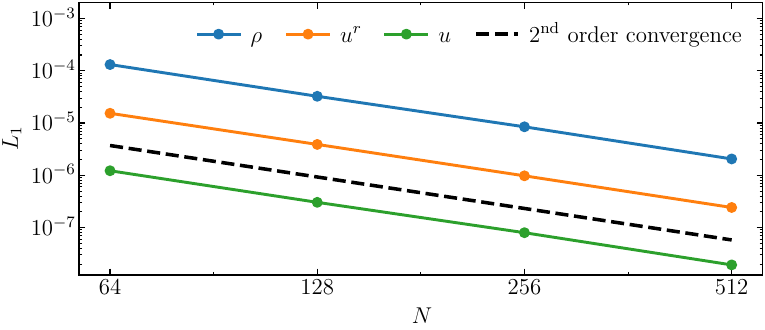}
	\caption{Convergence test for the self-gravitating polytropic star. The figure shows the $L_1$ errors of the primitive variables $\rho$, $u$, and $u^r$ as a function of the grid resolution $N$. The star is evolved without perturbations to test whether the coupling between the stress-energy tensor and the \mbox{self-gravity} solver preserves the initial hydrostatic equilibrium. The black dashed line indicates second-order convergence.}
	\label{fig:lane-emdem}
\end{figure}

We now test the coupling between hydrodynamics and self-gravity introduced through the source term in the energy--momentum
equation (Eq.~\ref{eqgrav}). We set the black hole mass and its spin to zero and initialize the analytical hydrostatic equilibrium profile of a star, given by the Lane--Emden polytropic model. Without self-gravity, such a star should start to expand due to pressure. We choose \mbox{$h=0.3$} for the coordinate transformation in MKS (Eq.~\ref{eq:MKS}), which additionally tests equilibrium in non-uniform coordinates. The HLLE Riemann solver with PPM reconstruction is chosen, and the Courant factor is set to \mbox{$\mathrm{CFL}=0.8$}.
For this unmagnetized test, we set $\varrho=\rho$ and keep only the rest-mass term in $\mathcal{I}^{\mu\nu}$ (Eq.~\ref{eq:mhd_inertia_tensor}), including the relativistic projection. For this test, we choose a polytropic index \mbox{$n=1$} for which the Lane-Emden equation simplifies to a linear differential equation. The
analytical formulae for the density and pressure are given by the following equations:
\begin{equation}
	\rho(r)
	=
	\rho_{\rm c}
	\frac{\sin(\pi r/R_\star)}{\pi r/R_\star},
	\qquad
	p(r)
	=
	\frac{2 R_\star^2}{\pi}\rho^2(r),
\end{equation}
where \mbox{$\rho_{\rm c}=10^{-4}$} is the central density and \mbox{$R_\star=10$} is the stellar radius. We test the numerical balance between pressure and gravity. We evolve the star without any perturbation and we measure the absolute $L_1$ errors of the density, internal energy, and radial velocity with respect to the initial state up to \mbox{$t = 10$}. The idea of this test is similar to the Fishbone--Moncrief stability test in Sect.~\ref{subsec:FM}. In Fig.~\ref{fig:lane-emdem} we show the absolute $L_1$ errors of the density, radial velocity, and internal energy. They decrease approximately at second order. We emphasize that this is not a Tolman--Oppenheimer--Volkoff test and does not model a fully relativistic star.

\subsection{Evrard collapse}
\label{evrard}
The final self-gravity test checks the energy conservation. In general, the implementations of self-gravity using source terms do not conserve the total energy with machine precision \citep{Jiang2013}. We use the Evrard collapse test \citep{Evrard1988} to check how our code handles the conversion of gravitational energy into kinetic and thermal energy. The density profile of a spherical Evrard cloud is given by:
\begin{equation}
	\rho(r) =
	\begin{cases}
		\dfrac{M_\star}{2\pi R_\star^2r}, & r\leq R_\star, \\[2mm]
		\rho_{\rm atm},                   & r>R_\star,
	\end{cases}
\end{equation}
where \mbox{$M_\star=R_\star=1$}, and the density of the atmosphere \mbox{$\rho_{\rm atm}=10^{-7}$}. The evolution of collapse is followed up to \mbox{$\tau=t\sqrt{\eta}/R_\star =3$}. The Poisson source is multiplied by \mbox{$\eta=10^{-4}$} to reduce the gravitational potential. The ideal gas equation of state is used with the adiabatic index \mbox{$\gamma_{\rm ad} =5/3$}. We use Minkowski spacetime, and reflective boundary conditions to prevent mass loss through the domain boundaries. The internal, kinetic, and gravitational potential energies are computed using non-relativistic approximations. We use a resolution of \mbox{$N_r\times N_{\theta}=1024\times 1024$}. We use the HLLE Riemann solver, with the PPM method for the reconstruction.

\begin{figure}
	\centering
	\includegraphics[width=0.99\linewidth]{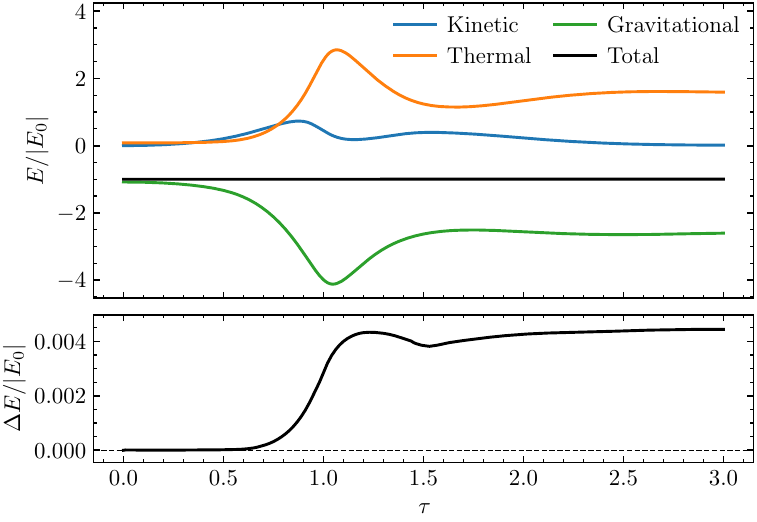}
	\caption{
		Energy conservation in the Evrard collapse test.
		The plot shows the gravitational, kinetic, thermal, and total energies over time. Because self-gravity is included through source terms, the total energy is not conserved to machine precision, but the error remains below one percent.
	}
	\label{fig:evrard}
\end{figure}

In Fig.~\ref{fig:evrard} we show the exchange of energy between the gravitational field, kinetic motion, and thermal energy during the collapse (top panel). The maximum error of the total energy conservation (bottom panel) is \mbox{$\Delta E /|E_0| \times 100\%\lesssim0.45\%$}. This value is comparable to that reported for a self-gravity module in the PLUTO code \citep{Mandal2023}. Therefore, we conclude that our code can conserve the total energy to better than one percent in the collapse scenario.

\section{Implementation and Performance}
\label{sect:perf}
\subsection{CUDA kernels and implementation}
{FUGUE} is a GPU-accelerated CUDA C++ code for high-performance calculations on NVIDIA GPUs. The code has a procedural, kernel-oriented CUDA design. The code supports multi-node communication using CUDA-aware MPI, with domain decomposition along all three spatial dimensions. Halo data are packed and unpacked by dedicated CUDA kernels. The core of the code is based on the CPU version of HARM \citep{HARM2003,HARM2006}. The metric-update method follows our previous works \citep{Janiuk2018,Dominika2021} and is implemented on GPUs. Additionally, we implement a weak-field self-gravity module using the spherical harmonic Poisson solver. Near the polar axis, the azimuthal cell size becomes very small, leading to very small timesteps. To avoid this, we optionally implement cylindrified coordinates {similar to those in HARMPI}\footnote{\url{https://github.com/atchekho/harmpi}}. {The cylindrification increases the cell size near the polar axis and allows larger timesteps.}

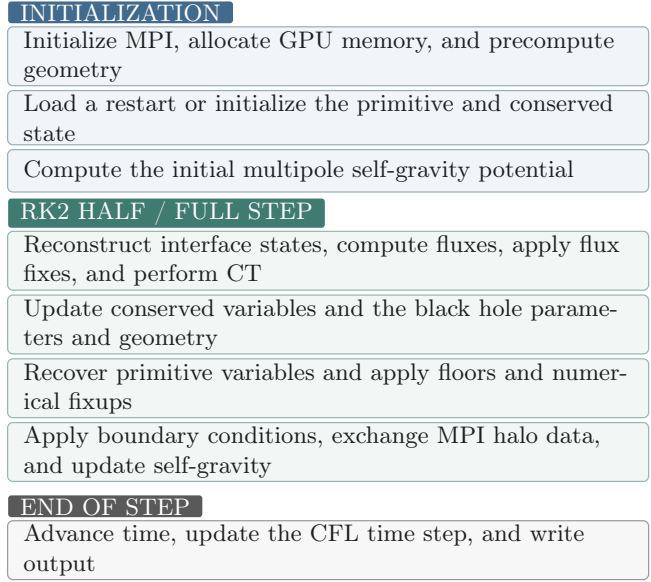
\begin{figure}[!htbp]
	\centering

	\begin{tikzpicture}[
			node distance=0.65mm,
			font=\normalfont,
			flowstep/.style={
					draw=black!35,
					line width=0.45pt,
					text=black!88,
					align=left,
					rounded corners=2pt,
					minimum height=5.6mm,
					text width=\dimexpr\linewidth-13pt\relax,
					inner xsep=6pt,
					inner ysep=1.6pt,
					outer sep=0pt
				},
			setup/.style={
					fill=setupaccent!7,
					draw=setupaccent!55
				},
			rk/.style={
					fill=halfaccent!7,
					draw=halfaccent!55
				},
			finish/.style={
					fill=black!3,
					draw=black!40
				},
			phasetag/.style={
					font=\normalfont,
					text=white,
					rounded corners=1.2pt,
					inner xsep=4.5pt,
					inner ysep=1pt
				}
		]

		\node[flowstep,setup] (k1)
		{Initialize MPI, allocate GPU memory, and precompute geometry};

		\node[flowstep,setup,below=of k1] (k2)
		{Load a restart or initialize the primitive and conserved state};

		\node[flowstep,setup,below=of k2] (k3)
		{Compute the initial multipole self-gravity potential};

		\node[flowstep,rk,below=5.2mm of k3] (k4)
		{Reconstruct interface states, compute fluxes, apply flux fixes, and perform CT};

		\node[flowstep,rk,below=of k4] (k5)
		{Update conserved variables and the black hole parameters and geometry};
		\node[flowstep,rk,below=of k5] (k6)
		{Recover primitive variables and apply floors and numerical fixups};

		\node[flowstep,rk,below=of k6] (k7)
		{Apply boundary conditions, exchange MPI halo data, and update self-gravity};

		\node[flowstep,finish,below=5.2mm of k7] (k8)
		{Advance time, update the CFL time step, and write output};

		\node[
			phasetag,
			fill=setupaccent,
			anchor=south west
		]
		at ([yshift=0.4mm]k1.north west)
		{{INITIALIZATION}};

		\node[
			phasetag,
			fill=halfaccent,
			anchor=south west
		]
		at ([yshift=0.4mm]k4.north west)
		{{RK2 HALF / FULL STEP}};

		\node[
			phasetag,
			fill=black!65,
			anchor=south west
		]
		at ([yshift=0.4mm]k8.north west)
		{{END OF STEP}};

	\end{tikzpicture}

	\caption{
		Simplified execution flow in {FUGUE}.
		The RK2 block is executed twice, first to construct the half-step
		state and then to perform the full-step update. Each stage includes
		interface reconstruction, flux calculation and correction,
		constrained transport (CT), conservative evolution,
		primitive-variable recovery, numerical fixups, boundary treatment,
		and MPI halo exchange. The Kerr metric evolution and multipole
		self-gravity operations are optional.
	}
	\label{fig:kernel_schedule}
\end{figure}

The schematic of the {FUGUE} kernel workflow is shown in Fig.~\ref{fig:kernel_schedule}. A standard single-domain GRMHD timestep involves more than 30 CUDA kernel launches, excluding MPI, self-gravity, and metric updates. The code implements the second-order Runge--Kutta (RK2) scheme for global evolution (Eq.~\ref{eq:RK2}), as well as time-dependent metric evolution (Sect.~\ref{sec:evol}) and the self-gravity module (Sect.~\ref{Sect:SG}). In the normal stationary spacetime scenario, the code calculates the geometry, e.g., metric components, connection coefficients, and determinant at the beginning. The code takes advantage of the fact that the Kerr metric is axisymmetric (Eq.~\ref{eq:Kerr}), and stores only one slice from three-dimensional grid in global memory. Because all geometric values are calculated on only one slice, the computational cost of refreshing geometry during metric update is small compared to the full three-dimensional GRMHD update.

When metric update is enabled, the code calculates and refreshes the geometry at every RK2 step (Eq.~\ref{eq:RK2}).
The metric update %
is made before recovering the primitive variables, to %
consistently adjust %
conserved variables (Eq.~\ref{eq:cons}) to the new geometry.
	{The HLLE/Lax--Friedrichs fluxes
		(Eq.~\ref{eq:HLLE} and Eq.~\ref{eq:lax}) at the inner boundary are used
		to determine the energy and angular-momentum accretion rates for metric update. In the code, we evolve the black hole mass and angular momentum using RK2, and recover the dimensional spin after each update.} This treatment correctly balances the conservation of energy and momentum in the simulation domain. The metric update works with either prograde \mbox{$(a>0)$} or retrograde \mbox{$(a<0)$} black hole spin configurations. The code can start with a Schwarzschild black hole and gradually spin it up, producing a rotating Kerr black hole.

We tested the summation of the energy and angular-momentum fluxes for all these configurations. The resulting changes in the black hole mass and angular momentum agree with the integrated inner-boundary fluxes to {round-off errors}. The quasi-stationary metric update was previously implemented and used in our CPU studies \citep{Janiuk2018,Dominika2021,Janiuk2023,Plonka2026}. Here, the implementation is different as we add a source term (Eq.~\ref{eq:S_t}) in the energy when the metric is updated. {The source term can be neglected, only when the fractional changes of the black hole mass and spin over a dynamical timescale are small.} To enhance the code's performance, we store slices of two derivatives (Eq.~\ref{eq:S_t_der}) to global memory, and the source term is added during update (Eq.~\ref{eq:update}), when the usual geometrical sources are added.

The Poisson solver is second order in time. The additional four-force (Eq.~\ref{eq:sg_source}) from self-gravity is applied during the update of conserved variables (Eq.~\ref{eq:update}). In the Poisson solver, the spherical harmonics and values associated with them are precomputed and saved in the global memory. Following the method described in \citet{Couch2013}, we precompute values of harmonics in the cell centers and on the cell interfaces. We read them at every iteration while making a projection of effective density onto the spherical harmonics (Eq.~\ref{eq:project}). Importantly, the entire Poisson solver operates in the MKS coordinates. We also tested the self-gravity module in magnetized flows using the full active source, and the complete passive response tensor. In these tests, the four-force remained orthogonal to the fluid four-velocity, and the evolution is numerically stable.

In {FUGUE}, we save the maximum values of the gravitational potential and divergence at intervals similar to those used for the mass accretion rate and other diagnostic parameters. We tested various scenarios, and we did not detect any problem in terms of divergence preservation, independently of the presence of the quasi-stationary evolving spacetime. For data output
we use the HDF5 format, and a single file is saved from all tasks in MPI processes.
The code also supports restarts. We have tested the code on several generations of NVIDIA GPUs, including Hopper, Ampere, Volta, and Kepler, showing the portability of the code across generations of GPUs.
\begin{figure}
	\centering
	\includegraphics[width=0.99\linewidth]{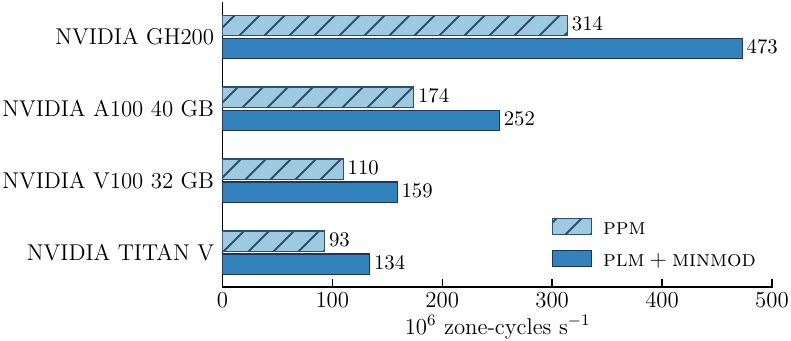}
	\caption{Zone-cycle performance for different single NVIDIA GPUs. The benchmark uses a
		$256^3$ Fishbone--Moncrief torus and RK2 time
		integration. The code solves the pure GRMHD equations, without self-gravity or
		metric evolution. The values are in units of $10^6$ zone-cycles s$^{-1}$.}
	\label{fig:zones}
\end{figure}
\subsection{Performance and scalability}
While developing {FUGUE}, we paid attention to performance. In contrast to the implementation presented by \cite{cuHARM}, we do not use shared memory, and as a goal, we minimize register usage to increase occupancy. The kernels are optimized to avoid register spilling. We use a structure-of-arrays (SoA) memory layout, which improves memory access on GPUs. When possible, we use 32-bit indices, making index calculations faster without changing the precision of physical variables. Another optimization is that the metric, inverse metric, metric determinant, and Christoffel symbols are precomputed before the simulation and stored in GPU memory. Spherical harmonics are also computed before the simulation. Minor improvements like the minimum CFL time step (Eq.~\ref{eq:timestep}) is first calculated inside each warp are implemented. Importantly, the code uses only double precision for the physical variables and for the Poisson solver.
In Table~\ref{tab:kernels} we present the relative kernel times for a three-dimensional simulation setup with the resolution \mbox{$256^3$} using Fishbone--Moncrief torus on an NVIDIA TITAN V. The solver is HLLE with the PPM method as the reconstruction scheme. The most demanding kernels are responsible for the flux calculations. The kernels for all three directions take about $50\%$ of the total computation time. The second most demanding kernel updates the conserved variables and includes all geometrical source terms, taking about $20\%$ of the total computation time. The primitive variable recovery kernel is next with about \mbox{$ 16\%$} relative cost.

	{We used NVIDIA Nsight Compute on an NVIDIA TITAN V to check how the main kernels use GPU compute resources and memory bandwidth. The average Compute (SM) Throughput and DRAM Throughput were $48\%$ and $27\%$ for flux calculation, $15\%$ and $69\%$ for the conserved variable update, and $46\%$ and $58\%$ for primitive recovery. These values may differ on other GPUs.}

In terms of performance, {FUGUE} achieves values similar to those of state-of-the-art codes like H-AMR \citep{H-AMR}. In Fig.~\ref{fig:zones}, we present the performance comparison (in zone-cycles $\rm s^{-1}$) for a few different NVIDIA GPUs, where one zone-cycle means one cell completes a full timestep, including both RK2 stages. We take into account only active cells. Using PLM reconstruction with the \textsc{MINMOD} limiter, our code achieves %
more than 470 million zone-cycles per second on a single NVIDIA GH200 GPU. The code with the Poisson solver set to \mbox{$\ell_{\rm max}=m_{\rm max}=16$} achieves %
156 million zone-cycles per second on an NVIDIA GH200, with PPM as the reconstruction method.

\begin{figure}
	\centering
	\includegraphics[width=0.99\linewidth]{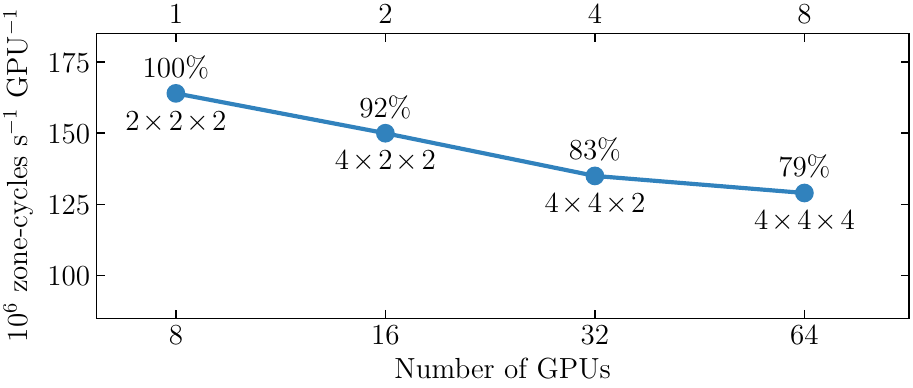}
	\caption{Weak scaling on NVIDIA A100 40GB GPUs, using a resolution of $256^3$ cells per GPU with a Fishbone--Moncrief torus. The top axis shows the number of nodes. The labels indicate the MPI decomposition. The efficiency remains about 80\% at 64 GPUs.}
	\label{fig:weak}
\end{figure}
\begin{table}
	\caption{\label{tab:kernel_timing}Relative CUDA kernel cost for the PPM reconstruction method with the RK2 time
		integrator on an NVIDIA TITAN V. The code uses the general-relativistic
		magnetohydrodynamics solver, without self-gravity or metric evolution. The
		resolution of the test is set to $256^3$, with a Fishbone--Moncrief torus.}
	\centering
	\setlength{\tabcolsep}{4pt}
	\renewcommand{\arraystretch}{1.12}
	\begin{tabular*}{\columnwidth}{@{\extracolsep{\fill}}lc}
		\specialrule{0.4pt}{0pt}{2pt}
		\specialrule{0.4pt}{0.3pt}{0pt}
		\noalign{\vskip 0.3mm}
		Kernel group & Time [\%] \\
		\hline
		Flux calculation & 50.7 \\
		Update with geometric source terms & 20.1 \\
		Primitive recovery and corrections & 15.8 \\
		Constrained-transport update & 7.9 \\
		Boundary and polar corrections & 1.2 \\
		Diagnostics and other kernels & 4.3 \\
		\specialrule{0.4pt}{0pt}{0pt}
	\end{tabular*}
	\label{tab:kernels}
\end{table}
We tested the MPI scalability of the code. In Fig.~\ref{fig:weak}, we present a weak scaling test on NVIDIA A100 40GB GPUs (8 cards per node). We use a resolution of $256^3$ cells per GPU with the magnetized Fishbone—Moncrief torus. The Riemann solver is chosen as the HLLE, with the PPM method as the reconstruction method. We achieved an efficiency of approximately $79\%$ for 64 GPUs across 8 nodes. We performed similar tests on 32 NVIDIA GH200 GPUs (4 cards per node) obtaining $81\%$ weak scaling efficiency in comparison to a single graphics card.

\subsection{Limitations}
\label{subs:Limitations}
The self-gravity module is a weak-field prescription on top of the Kerr background. This model is an effective weak-field approximation and is not a full solution of the Einstein equations. It implements a scalar Poisson solver, and applies a projected four-force. In the Poisson solver, we use auxiliary spherical coordinates, which introduces a spin-dependent error relative to oblate coordinates. The next numerically intented simplification is the usage of a flat Poisson solver (Eq.~\ref{eq:poi}). The solver in curved spacetime is far more computationally expensive. Another approximation is the instantaneous response of the self-gravity field. A potential solution to this problem is to evolve the self-gravity potential with a hyperbolic wave-equation solver \citep{Hirai2016}, which is beyond the scope of this work. Nevertheless, we note it as a potential improvement.

Importantly, it does not mean that the dynamics has the same relative errors. Near the black hole, the self-gravity source is much smaller than the geometrical momentum source, \mbox{$\chi=\left\lVert G_i \right\rVert /\left\lVert T^{\kappa}{}_{\lambda}\Gamma^{\lambda}{}_{i\kappa} \right\rVert\ll1$}; therefore, errors in the self-gravity force should have only a small local effect in this region. We note that the diagnostic $\chi$ is a coordinate- and slicing-dependent. For a test, we initialized the 16TI progenitor model \citep{Woosley2006,Urrutia2026} with a stellar mass of \mbox{$M_\star=13.95\,M_{\odot}$} and a black hole mass of \mbox{$M_{\rm BH}=3\,M_{\odot}$}. Using shell-integrated norms, for \mbox{$r<10\,r_g$}, we obtained \mbox{$\chi<10^{-3}$}. A small value of $\chi$ only shows that self-gravity is small compared with the geometric source near the black hole. We note that individual geometric, centrifugal, pressure or magnetic terms in the $\chi$ diagnostic can be partially canceled, so $\chi$ should be interpreted only as an approximate diagnostic. In addition to limitations, the multipole expansion is centered on the black hole, unlike in \citep{Couch2013}, and matter outside the domain is neglected.

	{ In our implementation, the surrounding matter can be more massive than the black hole if its gravitational potential is still weak, \mbox{$|\Phi|\ll1$}. However, if the matter distribution is strongly asymmetric, the black hole frame is not inertial frame of reference. This effect is included with the indirect potential from the \mbox{$\ell=1$} modes. It will studied in detail in the subsequent studies. The accuracy of the four-force prescription for relativistic flows remains to be tested against solutions with dynamical spacetime.}
\begin{figure*}
	\centering
	\includegraphics[width=0.999\linewidth]{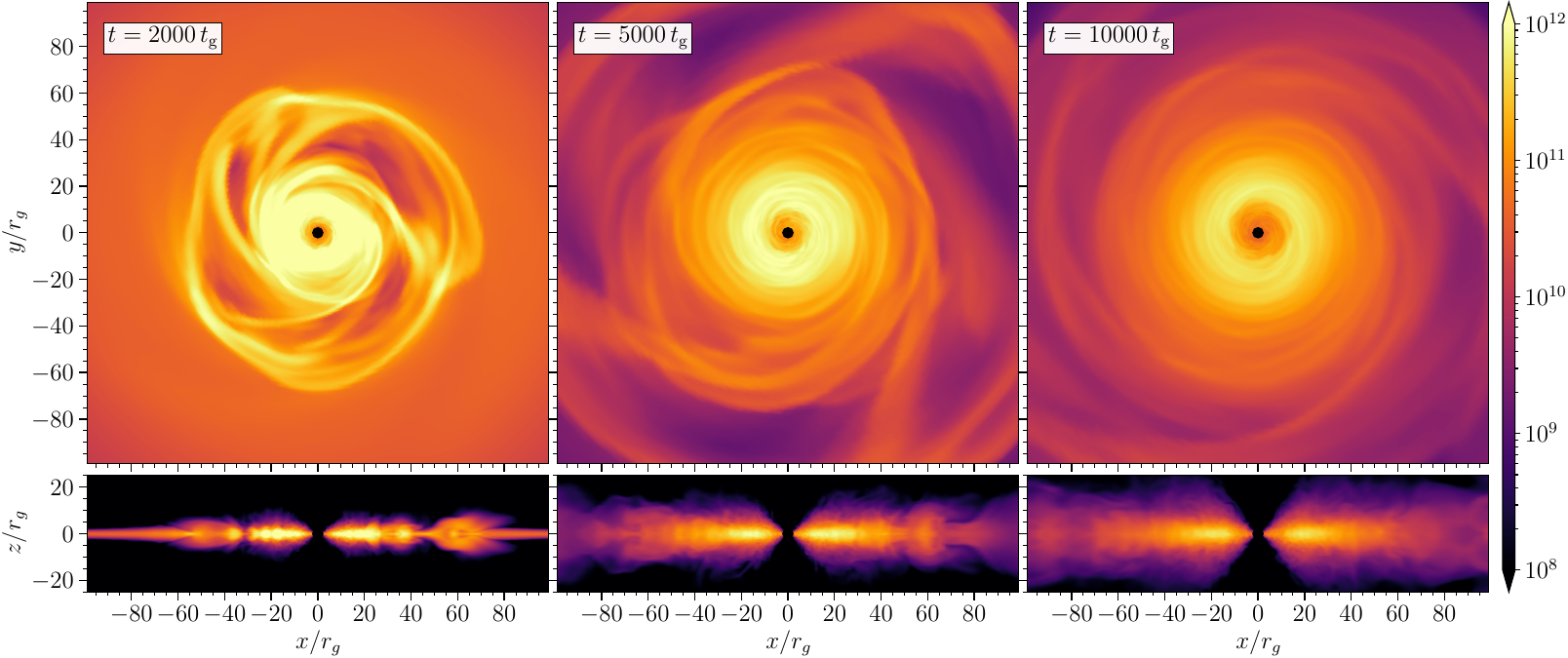}
	\caption{
	The maps of density $\rho\, (\mathrm{g\,cm^{-3}})$ for a Toomre unstable disk in the model including self-gravity. On the top panel, equatorial maps are shown, and on the bottom, poloidal slices are displayed. Over time, for \mbox{$t=2000\, t_g$}, \mbox{$t=5000\, t_g$}, and \mbox{$t=10000\, t_g$} the growth of non-axisymmetric structures is shown, starting from the regions with larger angular velocity in the vicinity of the black hole, and then to further regions. These figures show that {FUGUE} is able to track gravitational instabilities.}
	\label{fig:toomre_denisty}
\end{figure*}

\begin{figure}
	\centering
	\includegraphics[width=0.999\linewidth]{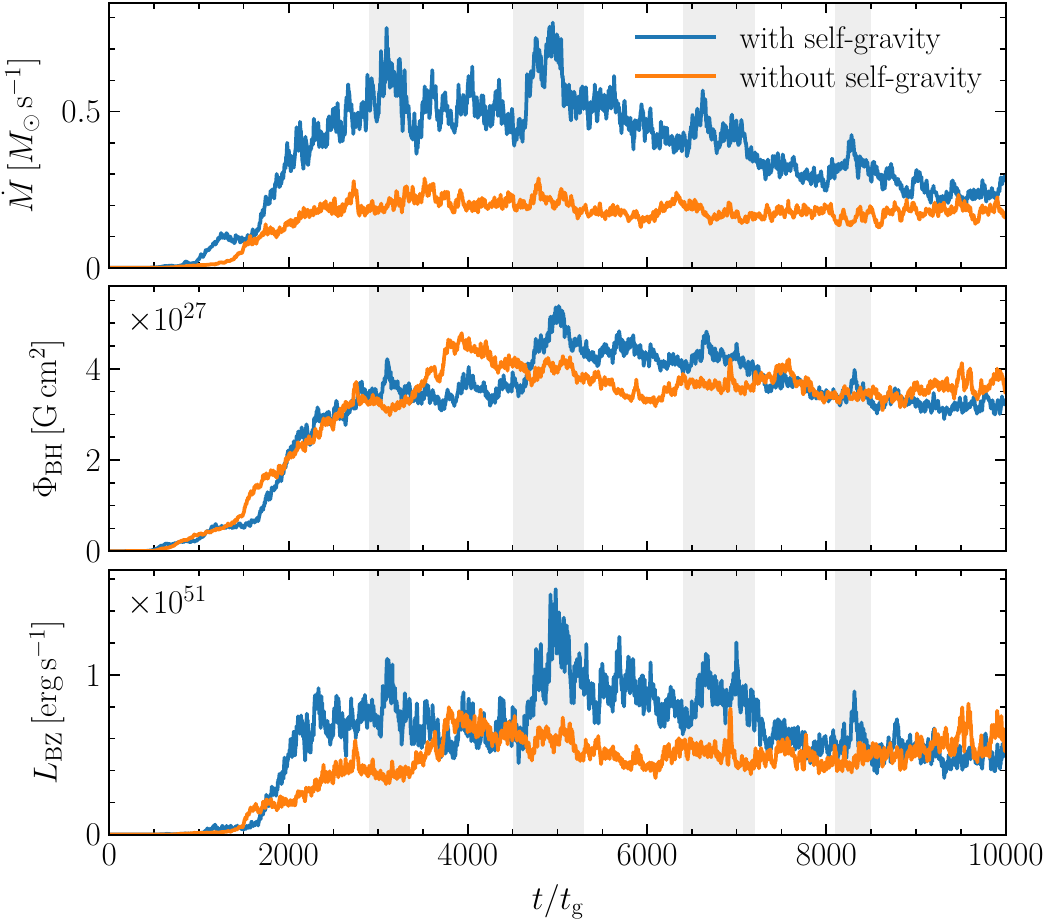}
	\caption{Time evolution of the mass accretion rate $\dot{M}$, magnetic flux threading the black hole $\Phi_{\rm BH}$, and Blandford--Znajek luminosity $L_{\rm BZ}$ for models with and without self-gravity. The blue and orange lines correspond to the self-gravitating and non-self-gravitating models, respectively. Periods of stronger accretion are followed by higher $L_{\rm BZ}$, suggesting that changes in the accretion flow may affect the jet power, potentially affecting prompt emission of GRBs.}
	\label{fig:toomre_evol}
\end{figure}
\section{{Toomre-unstable self-gravitating accretion disk}}
\label{sect:Toomre}
The development of the code is primarily motivated by collapsar simulations or massive accretion disks around black holes. Potentially, the collapsar neutrino-cooled disk can be gravitationally unstable \citep{Chen_2025}. The fixed-background GRMHD simulations \citep{Gottlieb2022} cannot capture gravitational instabilities. Simulations in dynamical spacetime can capture the growth of non-axisymmetric modes \citep{Korobkin2011,Wessel2023}; however, such simulations are extremely computationally expensive, which makes them nowadays impractical for long-term and large-scale simulations.

As a demonstrative model for FUGUE, we initialize two setups: one with self-gravity and one without. The mass of the accretion disk is set to \mbox{$M_{\mathrm{disk}}=1\, M_{\odot}$}, while the mass of the black hole is chosen to be \mbox{$M_{\mathrm{BH}}=3\, M_{\odot}$}, with the spin \mbox{$a_{\star}=0.8$}, and \mbox{$h=0.3$}. The disk scale height is set to a very thin value \mbox{$H/R=0.015$}. The disk has a Keplerian profile of angular velocity. The surface density is given by \mbox{$\Sigma\propto R^{-2}$}, and the vertical density profile is assumed to be Gaussian. The poloidal magnetic vector potential is set as \mbox{$A_\phi\propto \rho$}, and then the magnetic field is normalized to the plasma beta \mbox{$\beta = 100$} at \mbox{$r=12.5\, r_g$}. The outer radius is set to \mbox{$r_{\rm out}=300\, r_g$}. The equation of state is the ideal gas, with the adiabatic index $\gamma_{\rm ad}=4/3$. {For the adopted MKS mapping with \mbox{$h=0.3$}, the disk vertical thickness is resolved by approximately eight $\theta$-cells across $2H$ at \mbox{$r=12.5\,r_g$}. This demonstrative model is intended to capture global gravitational instabilities (GI) rather than fully resolved MRI turbulence.} We apply small perturbations to break symmetry in the hydrodynamic variables in order to seed gravitational instabilities. The disk is initially assumed to be Toomre unstable \citep{Toomre1964}, according to the criterion
\begin{equation}
	Q \equiv \frac{c_{\rm s}\kappa}{\pi G\Sigma},
	\label{eq}
\end{equation}
where $c_{\rm s}$ is the local sound speed, $\kappa$ is the epicyclic frequency, and $\Sigma$ is the disk surface density. We initialize the disk with a value \mbox{$Q\sim 1/3$}. We do not include any cooling source term during the evolution. Our disk is Toomre unstable initially, and over time, it can self-regulate by heating. The self-consistent effect of physical neutrino cooling will be the subject of further research \citep{Shibata2011,Radice2022}. The resolution of the simulations is set to \mbox{$(N_r\times N_\theta\times N_\phi)=(256\times256\times128)$}, and truncation of spherical harmonic solver set to \mbox{$\ell_{\rm max}=m_{\rm max}=16$}. We note that such values in the Poisson solver are too small for local disk fragmentation; nevertheless they can amplify global non-axisymmetric structures. The metric is updated quasi-stationarily during the evolution, as described in Sect.~\ref{sec:evol}. As a diagnostic, we use a several physical quantities. We calculate the mass accretion rate through the black hole event horizon:
\begin{equation}
	\dot{M}
	=
	-\left.
	\int_{0}^{2\pi}
	\int_{0}^{\pi}
	\rho u^{r}\sqrt{-g}\,
	d\theta\,d\phi
	\right|_{r=r_{\rm H}} .
	\label{eq:dotM}
\end{equation}
Next, we calculate the magnetic flux $\Phi_{\rm BH}$ threading the black hole given by:
\begin{equation}
	\Phi_{\rm BH}
	=
	\left.
	\frac{1}{2}
	\int_{0}^{2\pi}
	\int_{0}^{\pi}
	|B^{r}|\sqrt{-g}\,
	d\theta\,d\phi
	\right|_{r=r_{\rm H}},
	\label{eq:phiBH}
\end{equation}
where $B^r$ is the radial component of the magnetic field. We measure the electromagnetic luminosity by calculating the Blandford--Znajek power, given by:
\begin{equation}
	L_{\rm BZ}
	=
	-\left.
	\int_{0}^{2\pi}
	\int_{0}^{\pi}
	\left(T^{r}{}_{t}\right)_{\rm EM}
	\sqrt{-g}\,
	d\theta\,d\phi
	\right|_{r=r_{\rm H}} .
	\label{eq:L_BZ}
\end{equation}
We also define the volume-integrated amplitude of the $m$-th azimuthal Fourier modes \citep{2011PhRvL.106y1102K}, given by:
\begin{equation}
	D_m=\int_{r_{\rm H}}^{r_{\rm out}}
	\int_{0}^{\pi}
	\int_{0}^{2\pi}
	\rho u^t\,e^{- i m \phi}\,\sqrt{-g}\,
	d\phi\, d\theta\, dr.
\end{equation}

In Fig.~\ref{fig:toomre_denisty} we show the evolution of the Toomre unstable disk at three times: \mbox{$t=2000\, t_g$}, \mbox{$t=5000\, t_g$}, and \mbox{$t=10000\, t_g$}. The disk in the presence of self-gravity is expected to amplify non-axisymmetric modes or fragment from the innermost regions. The amplification is also visible in the poloidal slices, where inside the spiral arms the disk is thicker.  In Fig.~\ref{fig:toomre_evol}, we present the mass accretion rate $\dot M$ (Eq.~\ref{eq:dotM}), magnetic flux threading the black hole $\Phi_{\rm BH}$ (Eq.~\ref{eq:phiBH}), and the Blandford--Znajek luminosity $L_{\rm BZ}$ (Eq.~\ref{eq:L_BZ}). We note that for \mbox{$t<1750\,t_g$} the larger mass accretion rate \mbox{$\dot M\sim0.1\,M_\odot \,\rm s^{-1}$} is related to the changes in the %
effective potential. The disk is massive, and because the initial velocity was %
Keplerian, %
when self-gravity is included, the system needs to adjust to a new
distribution of the angular velocity which is %
changing
for an effective gravitational potential, %
given by the sum of black hole and disk masses. During this process, the %
magnetic flux is not accumulated. Then, after about \mbox{$t>2000\,t_g$}, the larger mass accretion rate is related to the episodic %
inflow pulses
due to disk instability, with strong episodic spikes, observed at
\mbox{$t\sim3000\, t_g$}, \mbox{$t\sim5000\, t_g$}, and \mbox{$t\sim7000\, t_g$}. An important %
effect of the model is that during episodic mass accretion, the magnetic flux on the event horizon is larger, and the Blandford--Znajek  luminosity grows as well. This is shown in Fig.~\ref{fig:toomre_evol} with gray shaded regions. It
suggests that in a physically consistent scenario, when the disk is unstable, during episodic mass accretion the electromagnetic power can be modulated, potentially affecting the variability of the jet and as a consequence gamma-ray burst emission properties. %
This effect will be the subject of our separate future studies.

To quantify the non-axisymmetric disk structure, we use the total normalized Fourier-mode power, given by:
\begin{equation}
	\mathcal{P}(t)
	=\sum_{m=1}^{16}\left|
	\frac{D_m(t)}{D_0(t)}
	\right|^2 ,
	\label{eq:mode_power}
\end{equation}
where the sum includes modes from \mbox{$m=1$} to \mbox{$m=16$}. In the model without self-gravity this quantity is approximately \mbox{$\mathcal{P}\sim10^{-4}$}, and the initial perturbations are damped. In the presence of self-gravity, they are amplified, and the total power is about a hundred times larger, with a value of \mbox{$\mathcal{P}\sim 10^{-2}$} throughout the simulation time. After \mbox{$t \sim 10\,000\,t_g$}, the disk gradually becomes gravitationally stable from the innermost regions due to progressive heating, and the differences between the model with self-gravity and the one without it start to disappear, as shown in Fig.~\ref{fig:toomre_evol}.

\begin{figure}
	\centering
	\includegraphics[width=0.999\linewidth]{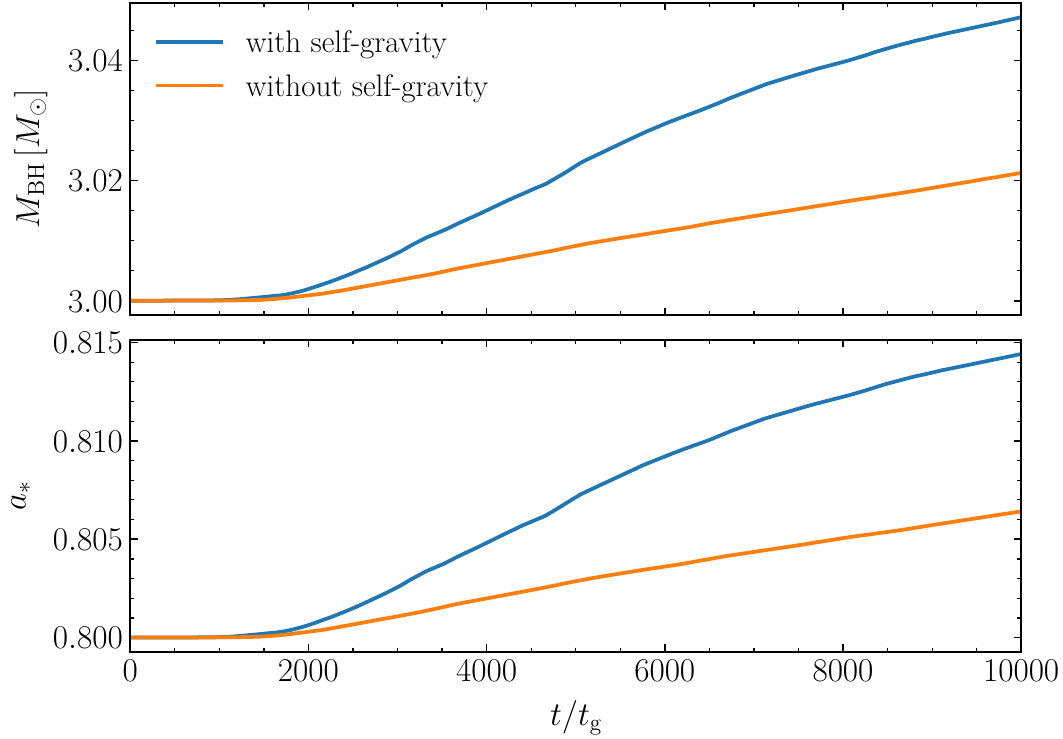}
	\caption{{Time evolution of black hole mass $M_{\rm BH}$ and dimensionless spin $a_{\ast}$ for models with and without self-gravity.}}
	\label{fig:toomre_aBH}
\end{figure}

{We present the evolution of the central black hole mass and dimensionless spin in Fig.~\ref{fig:toomre_aBH}. The accretion disk spins up the black hole, and the evolution of the spin and mass is highly correlated. Because of the larger mass and angular momentum accretion, the black hole spins up faster in the presence of self-gravity.} We note that throughout the evolution \mbox{$|\Phi_{\rm SG}|<10^{-2}$}, so the disk generates weak-field self-gravity even during the strong mode amplification. The indirect potential from the \mbox{$\ell=1$} coefficients was not included in
these simulations. Its effect \citep{Michael2010,Rafikov2026}, especially on the \mbox{$m=1$} mode and variability, will be studied
in future work.

\section{Conclusions}
\label{sect:Conclusions}
We have presented {FUGUE}, a new GPU-accelerated code for general relativistic magnetohydrodynamics (Sect.~\ref{sec:GRMHD}). The code implements an optional quasi-stationary Kerr black hole mass and spin evolution (Sect.~\ref{sec:evol}) and weak-field multipole self-gravity (Sect.~\ref{Sect:SG}). First and foremost, the code is a standard GRMHD code, optimized for HPC computing on NVIDIA GPUs. The performance of the code is comparable in terms of values of zone-cycles reported in other state-of-the-art GPU codes \citep{H-AMR,Asaf2024}, and because of the code's specialization, it is substantially faster than codes for dynamical spacetime, when those codes are used in a stationary spacetime scenario \citep{Shankar2023,AsterX}. The current implementation of {FUGUE} uses a fixed grid approach and does not include adaptive mesh refinement (AMR).

	{FUGUE} is a highly optimized code for solving the GRMHD equations. We validated the code across many numerical tests, including from shock tubes (Sect.~\ref{ST_REL} and Sect.~\ref{ST_MAG}),  Kelvin--Helmholtz instabilities (Sect.~\ref{KH}), through the magnetic rotator (Sect.~\ref{magrot}), blast wave (Sect.~\ref{blast}), and advective loop (Sect.~\ref{advection}). The general relativistic solver was validated with Michel--Bondi accretion (Sect.~\ref{bondi}) and the Fishbone--Moncrief torus (Sect.~\ref{subsec:FM}). {We compare results from FUGUE with state-of-the-art GRMHD codes (Sect.~\ref{sec:Porth}}). The Poisson solver was validated by testing its calculation of the gradient and potential for multipole and monopole scenarios (Sect.~\ref{App:monopole}). The convergence tests show second-order accuracy. As a demonstration scientific result, we showed that during the phase when the disk is gravitationally unstable, the mass accretion rate is more variable, and it also affects the Blandford--Znajek luminosity. These effects may influence the variability and emission properties of gamma-ray bursts; however, determining whether disks in collapsars are actually gravitationally unstable requires self-consistent simulations like \citet{2025ApJ...985L..26I} supplemented with self-gravity.

The code implements multipole self-gravity using a few simplifications. The main one is the weak-field requirement. However, after the formation of a black hole in collapsar simulations (e.g.,  \citealt{Urrutia2026}), the gravitational potential generated by the fluid is relatively weak. {The main approximations are the use of a flat Poisson solver, the prescribed self-gravity four-force, and the neglect of retardation effects.} However, in the regime for which the code is designed, these errors are the largest in the regions where the effects of self-gravity on dynamics are subdominant (Sect.~\ref{subs:Limitations}); nevertheless the dynamical effects depends on physical conditions.

The new code is primarily developed for our future collapsar studies. In long-duration 3D simulations, the effects of self-gravity can change the system evolution significantly \citep{Janiuk2023,Plonka2026}. Self-gravity reduces the dynamical timescale; hence, the collapse of a star proceeds faster. During a strong MAD state, the black hole spin-down should lead to an equilibrium within tens of seconds \citep{2024ApJ...961..212J,2025arXiv250208732I}. Our code, {FUGUE}, allows us to track the process of achieving spin equilibrium dynamically and very efficiently.

Another %
application of this code
is the formation of non-axisymmetric structures associated with gravitational instabilities, which can efficiently transport angular momentum and produce high, variable accretion rates \citep{Taylor2011}. The effects of mass accretion variability in the Toomre unstable disks were discussed in Sect.~\ref{sect:Toomre}, as a science case demonstrating the applicability of our code to new problems.
Such non-axisymmetric structures can be the source of gravitational waves \citep{Chen_2025, 2026MNRAS.545f1835L}.

Our plan for future code development
is to implement microphysics, such as that given by
the nuclear (Helmholtz) tabulated equation of state \citep{Janiuk2026}. We also plan to introduce the M1 method for neutrino transport \citep{Shibata2011,Radice2022,2025ApJ...985L..26I}.
These modifications are beyond the scope of the current article; however we highlight here the importance of nuclear physics and neutrino transport in the realistic collapsar simulations.

\begin{acknowledgements}
	We thank Ore Gottlieb and Yuri Levin for interesting discussions on the monopole/multipole self-gravity and Mateusz Kapusta for helpful suggestions on improving the performance of the code. {We thank the anonymous referee for helpful comments that improved our manuscript.} This work has been partially supported by grant No. 2023/50/A/ST9/00527 from the Polish National Science Center. We gratefully acknowledge Polish high-performance computing infrastructure PLGrid (HPC Center: ACK Cyfronet AGH) for providing computer facilities and support within computational grant no. PLG/2026/019543
\end{acknowledgements}

\appendix
\section{Magnetohydrodynamic inertial response tensor}
\label{TENSOR}
We start with the magnetohydrodynamic stress-energy tensor
(Eq.~\ref{eq:MHD}), which is given by
\begin{equation}
	T^{\mu\nu}=\left(w+b^2\right)u^\mu u^\nu+\left(p+\frac{b^2}{2}\right)g^{\mu\nu}-b^\mu b^\nu,
	\label{eq:appendix_Tmhd}
\end{equation}
where the enthalpy density is defined as
\mbox{$w\equiv\rho+u+p$}.
Taking the covariant divergence of Eq.~\ref{eq:appendix_Tmhd}, we obtain
\begin{equation}
	\begin{aligned}
		\nabla_\mu T^{\mu\nu}={} & u^\nu\nabla_\mu\left[\left(w+b^2\right)u^\mu\right]+\left(w+b^2\right)u^\mu\nabla_\mu u^\nu+\nabla^\nu\left(p+\frac{b^2}{2}\right)-b^\nu\nabla_\mu b^\mu-b^\mu\nabla_\mu b^\nu.
	\end{aligned}
	\label{eq:appendix_divT}
\end{equation}
We define the four-acceleration as \mbox{$a^\nu\equiv u^\mu\nabla_\mu u^\nu$}.
Contracting the ideal-MHD induction equation (Eq.~\ref{eq:GRMHD}) with $u_\nu$ and using
\mbox{$u_\mu u^\mu=-1$} and \mbox{$u_\mu b^\mu=0$}, we obtain the following relation between the divergence of the magnetic four-vector and the fluid four-acceleration, given by:
\begin{equation}
	\nabla_\mu b^\mu=b^\nu a_\nu.
	\label{eq:divb_acc}
\end{equation}
We project the energy-momentum equation onto the space orthogonal to the four-velocity using the projection tensor
\begin{equation}
	h^\alpha{}_\nu=\delta^\alpha{}_\nu+u^\alpha u_\nu,\qquad h^{\alpha\nu}=g^{\alpha\nu}+u^\alpha u^\nu.
	\label{eq:fluid_projector}
\end{equation}
Because \mbox{$h^\alpha{}_\nu u^\nu=0$}, the first term in Eq.~\ref{eq:appendix_divT} becomes zero after the projection. The four-acceleration is also perpendicular to the four-velocity, so \mbox{$u_\nu a^\nu=0$}. Therefore, \mbox{$h^\alpha{}_\nu a^\nu=a^\alpha$}.
The projected momentum equation is consequently
\begin{equation}
	\begin{aligned}
		h^\alpha{}_\nu\nabla_\mu T^{\mu\nu}={} & \left(w+b^2\right)a^\alpha+h^{\alpha\nu}\nabla_\nu\left(p+\frac{b^2}{2}\right)-b^\alpha\nabla_\mu b^\mu -h^\alpha{}_\nu b^\mu\nabla_\mu b^\nu.
	\end{aligned}
	\label{eq:projected_momentum}
\end{equation}
Using Eq.~\ref{eq:divb_acc} and the orthogonality relation
\mbox{$u^\nu a_\nu=0$}, so that
\mbox{$a^\alpha=(g^{\alpha\nu}+u^\alpha u^\nu)a_\nu$},
Eq.~\ref{eq:projected_momentum} can be written as
\begin{equation}
	\begin{aligned}
		h^\alpha{}_\nu\nabla_\mu T^{\mu\nu}={} & \underbrace{\left[\left(w+b^2\right)\left(g^{\alpha\nu}+u^\alpha u^\nu\right)-b^\alpha b^\nu\right]}_{\mathcal{I}^{\alpha\nu}}a_\nu+h^{\alpha\nu}\nabla_\nu\left(p+\frac{b^2}{2}\right)-h^\alpha{}_\nu b^\mu\nabla_\mu b^\nu.
	\end{aligned}
	\label{eq:projected_momentum_inertia}
\end{equation}
The tensor \mbox{$\mathcal{I}^{\alpha\nu}$} contains the terms related to the four-acceleration and is called the magnetohydrodynamic inertial response tensor. The other terms describe the gas and magnetic pressure gradients and the magnetic tension.
\bibliographystyle{aasjournalv7}
\bibliography{references}
\end{document}